# A General Framework to Forecast the Adoption of Novel Products: A Case of Autonomous Vehicles

**Subodh Dubey**[*]
Department of Transport and Planning
Delft University of Technology, Netherlands
s.k.dubey@tudelft.nl

**Ishant Sharma**[*]
Department of Civil Engineering
The University of Memphis, Tennessee, United States
isharma@memphis.edu

**Sabyasachee Mishra**
Department of Civil Engineering
The University of Memphis, Tennessee, United States
smishra3@memphis.edu

**Oded Cats**
Department of Transport and Planning
Delft University of Technology, Netherlands
o.cats@tudelft.nl

**Prateek Bansal**[†]
Department of Civil and Environmental Engineering
National University of Singapore, Singapore
ceev236@nus.edu.sg

[*]Authors contributed equally to this work.
[†]Corresponding Author.

**Abstract**

Due to the unavailability of prototypes, the early adopters of novel products actively seek information from multiple sources (e.g., media and social networks) to minimize the potential risk. The existing behavior models not only fail to capture the information propagation within the individual's social network, but also they do not incorporate the impact of such word-of-mouth (WOM) dissemination on the consumer's risk preferences. Moreover, even cutting-edge forecasting models rely on crude/synthetic consumer behavior models. We propose a general framework to forecast the adoption of novel products by developing a new consumer behavior model and integrating it into a population-level agent-based model. Specifically, we extend the hybrid choice model to estimate consumer behavior, which incorporates social network effects and interplay between WOM and risk aversion. The calibrated consumer behavior model and synthetic population are passed through the agent-based model for forecasting the product market share. We apply the proposed framework to forecast the adoption of autonomous vehicles (AVs) in Nashville, USA. The consumer behavior model is calibrated with a stated preference survey data of 1,495 Nashville residents. The output of the agent-based model provides the effect of the purchase price, post-purchase satisfaction, and safety measures/regulations on the forecasted AV market share. With an annual AV price reduction of 5% at the initial purchase price of $40,000 and 90% of satisfied adopters, AVs are forecasted to attain around 85% market share in thirty years. These findings are crucial for policymakers to develop infrastructure plans and manufacturers to conduct an after-sales cost-benefit analysis.

**Keywords:** word of mouth; risk aversion; novel technology adoption; social network; autonomous vehicles.

# 1. Introduction

Capturing consumers' intention to purchase or adopt a novel product is vital for multiple disciplines such as transportation, marketing, sales, technology, economics, finance, human-machine interaction, and social behavior, among others. We contribute to this interdisciplinary literature by providing a general framework to elicit consumers' preferences and forecasting their adoption of "really new products" (i.e., innovative products with entirely new or different attributes from any existing products) (Gregan-Paxton and John, 1997). Autonomous vehicle (AV) – a fully-automated self-driving privately-owned vehicle – is a case in point, which falls under this product category with recent innovations in technology-assisted motorized driving. Understanding consumer preferences and forecasting adoption rates of AVs are crucial for policymakers to devise a plan to meet infrastructure needs and make regulatory decisions to manage a mixed fleet of AVs and conventional vehicles (CVs). At the same time, quantifying the effect of the purchase price and marketing strategies on AV adoption rate is equally critical for manufacturers to conduct an after-sales cost-benefit analysis.

## 1.1. Background and Motivation

Potential consumers can state their intentions to adopt an existing product based on the attributes of interest, by means of trial periods and test drives. In contrast, capturing consumer's intention to purchase a futuristic product is challenging due to the unavailability of accessible prototypes for first-hand experience. Therefore, early adopters actively search for information about the anticipated attributes, benefits, and barriers associated with novel products to minimize associated risks and maximize post-purchase satisfaction (Dholakia, 2001; Dowling and Staelin, 1994; Liu, 2013; Manning et al., 1995; Mosley and Verschoor, 2005).

With recent technological advancements in smartphones and ubiquitous internet connectivity, potential consumers are increasingly exchanging their opinions and recommendations about innovations through electronic media, social media, blogs, and peer-to-peer communication, among other communication channels and informational sources (Gupta and Harris, 2010; Ha, 2002). The information obtained from such channels is commonly referred to as *word-of-mouth* (WOM). In this era of a digital revolution, the influence of the product-related information through WOM on consumer preferences has become substantial enough to be carefully accounted for in econometric models (Huete-Alcocer, 2017). Due to the inability to experience the product, WOM plays an even more vital role in alleviating or increasing risks associated with the adoption of novel product (Hirunyawipada and Paswan, 2006; Hussain et al., 2018, 2017; Krishnamurthy, 2001; Manning et al., 1995; Tan, 1999). Such differences in product information transmission and its effect on the consumers' risk perception call for specific consumer behavior models for novel products.

We envision that an ideal econometric model to elicit the consumers' preferences for novel products should have five components: consumer's risk preferences, WOM through offline social networks and online channels, the interplay between consumer's risk preferences and WOM, adoption level of the product in social network/city, and the influence of key product attributes such as purchase price on consumer's preferences. Followed by generating a synthetic population, such a consumer preference model can be integrated into an agent-based model to forecast the adoption of the novel product under different scenarios (e.g., purchase price reduction or changes in risk preferences due to technological improvements such as reduction in AV crash rate).

**1.2. Research Gap**

The literature on capturing the impact of WOM on consumer preferences is prolific (see Table A.1 in supplementary material for a summary of 40+ such studies). Specific to the transportation sector, structural equation models (Kwon et al., 2020; Thøgersen and Ebsen, 2019), discrete choice models (He et al., 2014; Helveston et al., 2015; Jansson et al., 2017), agent-based models (Kieckhäfer et al., 2017), exploratory factor analysis (Ozaki and Sevastyanova, 2011), regression analysis (Barth et al., 2016; Du et al., 2018; Moons and De Pelsmacker, 2012), text mining (Ma et al., 2019), theory of reasoned action (Alzahrani et al., 2019), and Bass model (Hong et al., 2020) have been used to model WOM and social network effects in the adoption of green or electric vehicles (EVs). Specific to AVs, Ghasri and Vij (2021) explored the influence of WOM on the consumer preferences using discrete choice. Most past studies could only quantify consumers' preferences to adopt EVs or AVs, but failed to translate the estimated preferences into a forecasting model. For instance, the results of structural equation models on *consumers' intention to purchase* are not adequate for forecasting the adoption of novel products. Only He et al. (2014) and Kieckhäfer et al. (2017) extended the analysis to incorporate the effect of WOM in forecasting the adoption of EVs, and Talebian and Mishra (2018) did the same for AVs. However, these studies lack the underlying consumer behaviour model.

Similarly, only a handful of previous studies used structural equation models (Chikaraishi et al., 2020), the technology acceptance model (Zhang et al., 2019), or discrete choice models (Bansal et al., 2021; Wang and Zhao, 2019) to understand consumers' risk preferences in the adoption of EVs and AVs. However, none of these studies forecasted the market penetration of new technologies. A similar pattern was observed in modeling risk preferences for other novel products such as farming techniques (Barham et al., 2014; Brick and Visser, 2015). To substantiate our claim, we summarize the related literature in Table A.1 in supplementary material. There are also many studies in the literature, which completely ignored WOM and risk preferences, and relying exclusively on product attributes (such as purchase price) in agent-based models to forecast the adoption of novel products such as EVs and AVs (e.g. Bansal and Kockelman, 2017; Musti and Kockelman, 2011).

In summary, numerous studies modeled WOM and risk preferences in eliciting consumers' inclination towards novel products, but they have three main shortcomings. First, previous studies fail to explicitly incorporate the social network effect. A handful of studies have modeled the effect of information from internal or external sources on consumer preferences (for example, Ghasri and Vij, 2021; Sharma and Mishra, 2020) but have failed to model *information propagation* within the social network. Therefore, these consumer behavior models cannot be used to forecast the adoption of novel products. Second, none of the previous studies has simultaneously accounted for the effect of WOM and risk aversion on consumer behavior. Third, while only a handful of studies have gone beyond consumer behavior analysis and forecasted the adoption of the novel product, forecasting models rely on simplistic (mostly synthetic, i.e., not calibrated with the contextual data) consumer behavior models.

### 1.3. Contributions

We propose a comprehensive framework to forecast the adoption of the novel product where a well-calibrated new consumer behavior model is integrated into a population-based agent-based model. Specifically, the consumer behavior model follows the specification of the integrated choice and latent variable (ICLV) model (also known as the hybrid choice model), where WOM and risk aversion are considered as latent variables. The proposed model – interdependent ICLV – extends the ICLV model (Bhat et al., 2016b) by incorporating cross-loading of latent variables and panel effects in the discrete choice component. In this specification, autoregressive structure in the latent construct of ICLV captures social network effects, and cross-loading WOM on risk aversion accounts for their interaction effects. The indirect utility of the choice model captures the effect of the purchase price and adoption rates of the product within the social network and the city. We derive the maximum likelihood estimator of the interdependent ICLV model and calibrate it with stated preference data. The calibrated consumer behavior model, synthetic population, and individual-level social network are passed through an agent-based simulation model to predict individual's preferences to buy the novel product in each time step. The individual-level preferences are aggregated to find the adoption rate in each time step. We also highlight the implications of neglecting the information propagation effect in the agent-based simulation. The proposed framework is general enough to be applied for forecasting the adoption of any novel product. However, to make the discussion contextual, we demonstrate its capabilities in forecasting the adoption of AVs.

      The remainder of the paper is organized as follows: Section 2 discusses the design of stated choice experiments to collect the preferences of Nashville (Tennessee, USA) residents for AV adoption. The details of data collection and summary statistics are also presented in this section. Section 3 provides specific details of the considered aspects of WOM and risk preferences, followed by a contextual and mathematical representation of the interdependent ICLV model. Section 4 summarizes the results of the ICLV model. Section 5 details the synthetic population generation, the agent-based simulation framework, and scenario-based analysis of the forecasted adoption rates of AVs in Nashville. Conclusions and avenues of future research are discussed in the final section.

## 2. Survey Design and Data Collection

### 2.1 Discrete Choice Experiment Design

We designed and conducted a stated preference survey with a discrete choice experiment (DCE). In the DCE, respondents were asked to choose between a CV and an AV during their next car purchase. They needed to make choices based on the purchase prices of both cars and the adoption of AVs in their social network and city. Before the experiment, respondents were informed about the differences between AVs and CVs using infographics. To ensure that respondents do not find automation very futuristic, we also mentioned that Google's AV has driven more than 20 million miles on public roads. We communicated that Level-5 AVs are slightly more expensive than CVs because they need additional accessories to operate without a human driver, but both cars are equivalent in all other attributes (e.g., engine power, fuel economy, body type, and aesthetics). Attribute levels of both alternatives in the DCE are presented in Table 1. The purchase price of AV was pivoted on the purchase price of the CV, which was asked in a question preceding the DCE.

Four attribute levels for the purchase price and three attribute levels for each social network and city level AV adoption lead to a total of thirty-six choice scenarios. We used the full factorial design and presented each respondent with three randomly selected scenarios out of thirty-six choice scenarios. An example of the choice scenario presented to respondents is shown in Figure 1.

**Table 1: Attribute levels in the discrete choice experiment and experiment design to capture word of mouth (WOM) effect related to the adoption of autonomous cars**

| Attribute | Alternatives | |
|---|---|---|
| | **Conventional car** 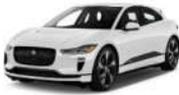 | **Autonomous car** 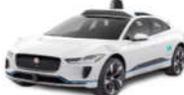 |
| **Discrete choice experiment** | | |
| Purchase price (US$) | Reported by the respondent | 1. 20% higher than the cost of conventional car<br>2. 30% higher than the cost of conventional car<br>3. 40% higher than the cost of conventional car<br>4. 50% higher than the cost of conventional car |
| % of people in **respondent's social network** who adopted autonomous cars | Not applicable | 1. 30%<br>2. 60%<br>3. 90% |
| % of people in **respondent's city** who adopted autonomous cars | Not applicable | 1. 30%<br>2. 60%<br>3. 90% |
| **WOM Experiment 1: safety aspects of autonomous cars** | | |
| Source of information | 1. Friend<br>2. Car dealer<br>3. Colleague<br>4. Media | Same as the one for conventional car |
| Vehicle crashes per 100 million miles | 1,090 | 1. 415<br>2. 290 (30% less than 415)<br>3. 207 (50% less than 415) |
| Crashes with no clarity on responsibility/liability | Not applicable | 1. 10%<br>2. 30% |
| **WOM Experiment 2: environmental friendliness, travel time savings, and safety aspects of autonomous cars** | | |
| Source of information | 1. Friend<br>2. Car dealer<br>3. Colleague<br>4. Media | Same as the one for conventional car |
| Travel time reduction in autonomous cars | Not applicable | 1. 20% less than conventional cars<br>2. 40% less than conventional cars |
| $CO_2$ emissions reduction in autonomous cars | Not applicable | 1. 30% less than conventional cars<br>2. 50% less than conventional cars |
| Crashes with no clarity on responsibility/liability | Not applicable | 1. 10%<br>2. 30% |

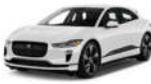

**Figure 1: An example of a choice situation presented to respondents.**

## 2.2 WOM Experiments

Apart from the DCE, the survey also had two experiments to capture the type and magnitude of the WOM transmitted by respondents based on the source of information and AV attributes. In both experiments, we used four information sources: friend, car dealer, colleague, and media. All the attribute levels of both experiments are tabulated in **Error! Not a valid bookmark self-reference.**.

In experiment 1, we provided information about vehicle crashes and fatality rates for CVs and AVs, and the percentage of AV crashes with no clarity about who is responsible for the crash. Following Blanco et al. (2016), we used 1,090 and 415 vehicle crashes per 100 million miles as crash rates for CVs and AVs, respectively. Assuming a potential reduction of 30-50% in the number of AV crashes, we considered 415, 290, and 207 as the three levels of the AV crash rate. Whereas we used 1.13 fatalities per 100 million miles for CVs (IIHS, 2020), it was considered to be zero for AVs because Google's AV did not report any fatality in the year 2018 and 2019 (Waymo, 2020). We also assumed that making anyone accountable or responsible for the crash might be challenging in 10-30% of AV crashes. Therefore, we considered 10% and 30% as two levels for this attribute. Experiment 2 is similar to experiment 1, but crash rates and fatalities were replaced with reductions in travel time and $CO_2$ emissions due to automation. The past findings suggest that AVs are expected to reduce travel time and $CO_2$ emissions by 37% and 30%, respectively, at 50% market penetration (Olia et al., 2016). Based on this information, we used attribute levels of 20% and 40% for travel time reduction and 30% and 50% for $CO_2$ emission reduction.

The full factorial designs of experiments 1 and 2 have (4x3x2) 24 and (4x2x2x2) 32 choice scenarios. One randomly selected scenario was presented to each respondent for one of the two experiments. Readers will note that both experiments ensure a trade-off between the benefits of automation (crash/fatality reduction in experiment 1, and travel time and emissions reduction in experiment 2) and the associated risks (AV crashes with no clarity on responsibility). Considering these trade-offs, the respondent was asked to rate three statements, each corresponding to positive, neutral, and negative WOM, on a five-point Likert scale (from strongly disagree to strongly agree) in each experiment. Specifically, based on the presented information in the experiment, the respondents were asked how likely they would positively, neutrally, or negatively communicate their opinion about AV adoption to their close social ties. The narratives of both experiments 1 and 2, along with the WOM statements, are shown in example scenarios presented in Figure 2 and Figure 3, respectively.

Suppose **your friend** provides you with the following information on the **crash and fatality rates** of conventional and autonomous cars if both vehicles are driven for **100 million miles:**

| Conventional cars | | Autonomous cars | |
|---|---|---|---|
| Crashes | Fatalities | Crashes | Fatalities |
| 1,090 | 1.13 | 290 | 0.00 |

The above table indicates that autonomous cars have a much lower crash and fatality rates, but **your friend** also informs that it is not clear who is responsible for the crash in **30%** of crashes encountered by **autonomous cars**.

Assume people in your close social network ask about your opinion on buying an **autonomous car over a conventional car.** Based on the information presented above, t**o what extent do you agree the following statements:**

|  | Strongly agree | Somewhat agree | Neutral | Somewhat disagree | Strongly disagree |
|---|---|---|---|---|---|
| I will suggest them to **consider buying an autonomous car** over a conventional car because the former is much **safer**. | ○ | ○ | ○ | ○ | ○ |
| I will be **neutral** in my recommendation. | ○ | ○ | ○ | ○ | ○ |
| I will suggest them to **consider buying a conventional car** over an autonomous car because at least one knows who is **responsible for a crash** in a conventional car. | ○ | ○ | ○ | ○ | ○ |

**Figure 2: An example scenario that captures WOM related to the safety of autonomous cars (experiment 1).**

> Suppose **a print or electronic media channel** provides you with the following information on the travel time and emission reduction in autonomous cars as compared to conventional cars if both vehicles have **an equal market share** (i.e., 50% market penetration of autonomous cars):
>
> | Autonomous cars | |
> |---|---|
> | Reduction in travel time | Reduction in $CO_2$ emissions |
> | 20% | 30% |
>
> But **the media channel** also informs you that there are two potential limitations of autonomous cars:
>
> 1. Whereas you can accelerate the conventional car to reach on time for important meetings or flights, you will not have such flexibility to make the last moment decision while riding autonomous cars.
> 2. During an accident, about **30%** of the time it will not be clear who is responsible for the crash encountered by autonomous cars.
>
> Assume people in your close social network ask about your opinion on buying an **autonomous car over a conventional car**. Based on information presented above, **to what extent do you agree to the following statements:**
>
> |  | Strongly agree | Somewhat agree | Neutral | Somewhat disagree | Strongly disagree |
> |---|---|---|---|---|---|
> | I will suggest them to **consider buying an autonomous car** over a conventional car because the former is more **reliable and environmental-friendly**. | ○ | ○ | ○ | ○ | ○ |
> | I will be **neutral** in my recommendation. | ○ | ○ | ○ | ○ | ○ |
> | I will suggest them to **consider buying a conventional car** over an autonomous car because at least one can **make last-moment decisions to accelerate and who is responsible for a crash** in a conventional car. | ○ | ○ | ○ | ○ | ○ |

**Figure 3: An example scenario that captures WOM related to environmental friendliness, travel time savings, and safety aspects of autonomous cars (experiment 2).**

### 2.3 Data Collection

The web-based stated preference survey was hosted on Qualtrics and was disseminated among Nashville residents between August and November 2020. Survey participants were recruited with the help of an online market research firm. The participants were asked screening questions regarding age and the city of residence to ensure that only Nashville residents with age over 18 years participate in the survey. The respondents were also asked about the five-digit ZIP code of the home location, which was subsequently used to generate synthetic population and social networks (see Sections 5.1 and 3.3.6 for details). To ask for such detailed residential information, we had to take approval from the Institutional Review Board (IRB) at the University of Memphis under the "*Expedited*" track. The Zipcode-level spatial distribution of the respondents in Nashville is displayed in Figure 4.

Apart from the DCE and two WOM experiments, the survey asked respondents about their socioeconomic characteristics, accident history, and social ties (see Figure A.1 in the supplementary material for the definition of a close social tie). We also asked respondents to report their perceptions about independent statements on the five-point Likert scale – the indicators of the respondent's anticipated risk in adopting AVs. To maintain the sample quality, participants with a response time below 50% of the median response time were removed. The final sample had 1,495 complete responses. The marginal distributions of sample and population (as per American Community Survey 2017 obtained from Manson et al., 2019) across ethnicity, gender, and age are shown in Figure 5. The overall sampling distribution across all demographic levels is comparable to the population distribution.

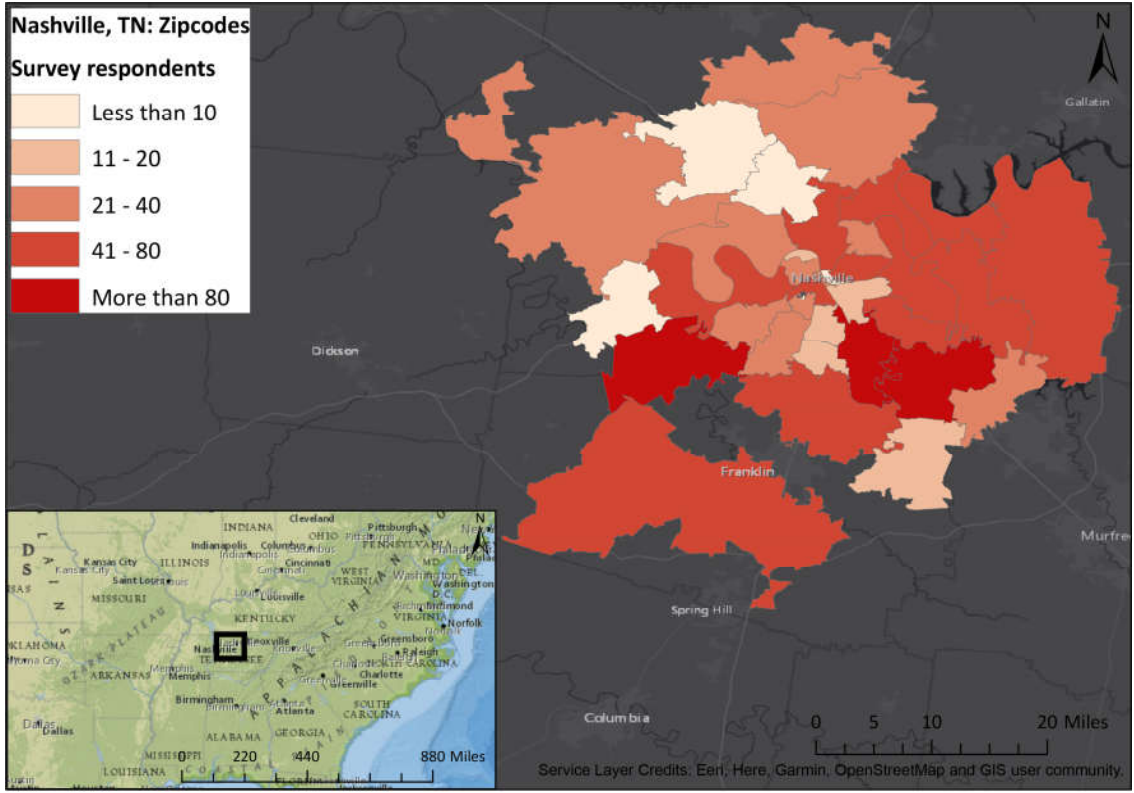

**Figure 4: Spatial distribution of survey respondents in Nashville**

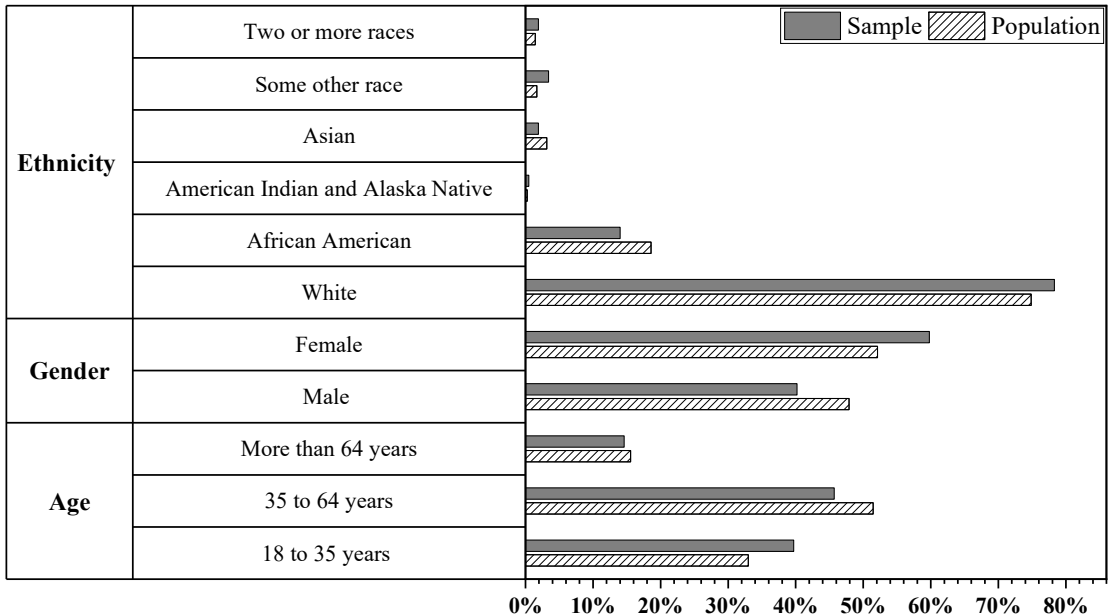

**Figure 5: Marginal distribution of sample and population across demographics.**

## 2.4 Summary Statistics

Table 2: summarizes the descriptive statistics of the survey sample. From a social network perspective, respondents have about 12 close social ties. Among the individuals with accident history, most of them incurred minor damages, and a minority of the sample (6%) suffered from severe injuries.

The distributions of responses to the statements indicating the risk perception of the respondent on the five-point Likert scale are shown in Figure 6. Over half of respondents are worried about the value-for-money in an AV purchase (Ind01). Only 27% of the respondents are willing to take a risk of purchasing an AV to have an exciting experience (Ind02). 58% of the respondents indicated that they would be uncomfortable in switching to AVs (Ind03), but this might not be specific to automation technology because around 45% of respondents generally struggle with such risky decisions (Ind04).

**Table 2: Descriptive statistics of the sample**

| Categorical variables | | | |
|---|---|---|---|
| **Variable** | **Percentage** | **Variable** | **Percentage** |
| Gender | | Any kind disability undermining ability to drive | |
| Male | 40% | Yes | 11% |
| Female | 60% | No | 89% |
| Age | | Ethnicity | |
| 18 to 35 years | 40% | White | 78% |
| 35 to 65 years | 46% | African American | 14% |
| more than 65 years | 15% | Others | 8% |
| Educational Attainment | | Annual household income | |
| High school or below | 20% | less than $25,000 | 16% |
| Some College or College graduate | 61% | $25,000-$35,000 | 11% |
| Master's (MS) or Doctoral Degree (Ph.D.) | 15% | $35,000-$75,000 | 35% |
| Professional Degree (MD, JD, etc.) | 4% | $75,000-$125,000 | 22% |
| | | More than $125,000 | 16% |
| Willingness to pay to purchase a new car | | Number of children in household | |
| less than $15,000 | 31% | Zero | 65% |
| $15,000-$30,000 | 40% | One or more | 35% |
| more than $30,000 | 29% | | |
| Involve in accidents where vehicle incurred minor damages | | Involved in accidents where vehicle incurred major damages | |
| Yes | 56% | Yes | 36% |
| No | 44% | No | 64% |
| Involved in accidents and suffered from minor injuries | | Involved in accidents and suffered from severe injuries | |
| Yes | 24% | Yes | 6% |
| No | 76% | No | 94% |
| Continuous variables | | | |
| Number of vehicles in the household | | Number of workers in the household | |
| Mean | 2.82 | Mean | 1.58 |
| Standard deviation | 0.93 | Standard deviation | 0.66 |
| Household members | | Number of social ties | |
| Mean | 2.53 | Mean | 11.66 |
| Standard deviation | 1.03 | Standard deviation | 61.22 |

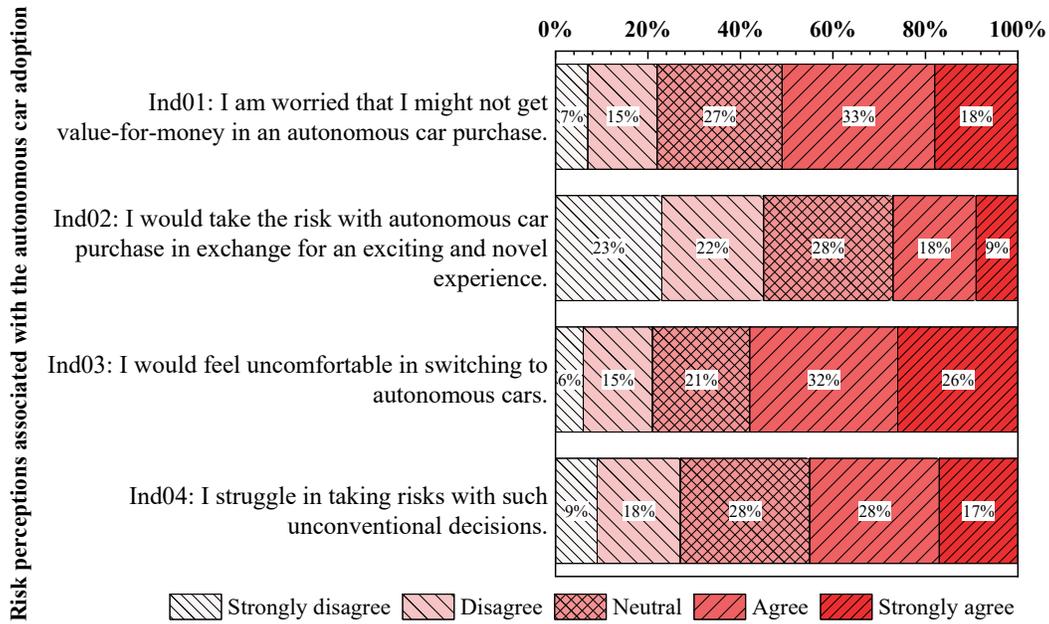

**Figure 6: Descriptive statistics of statements related to risk perception associated with the autonomous car adoption.**

## 3. Consumer Behavior Model

### 3.1 Word of Mouth (WOM)

There is well-documented evidence of social influence in the purchase of ice cream (Richards et al., 2014), electronic equipment (Narayan et al., 2011), smartphone (Park and Chen, 2007), organic food items (Chen, 2007), and automobile (Grinblatt et al., 2008). The literature classifies social influence into three major outcomes – conformity, compliance, and obedience (Maness et al., 2015). Conformity is the most common social influence that occurs when an individual changes the behavior to gain acceptance in a group or improve social status by impressing others. Compliance and obedience occur in situations where an individual is *requested* and *ordered* to change the behavior, respectively. Conformity is the most plausible form of social influence in the context of AVs and other novel products. Our review suggests that there could be four ways to model information transfer through WOM and conformity behavior.

The *first* approach is based on the threshold-based specification, which defines a threshold on the proportion of population and friends who must adopt the product before the individual does so (Granovetter, 1978). Such effects need to be incorporated in AV adoption because previous studies on EVs have found the presence of threshold effects. For instance, Mau et al. (2008) observed a significant increase in the individual's willingness to pay for EVs as a function of overall market share. Threshold effects can be incorporated by explicitly modeling a willingness to adopt innovation as a function of market share (Zhang et al., 2011), or an additive utility component can be added based on the choice of others in the network (He et al., 2014; Hsu et al., 2013; Kim et al., 2014; Rasouli and Timmermans, 2016). The *second* approach considers that the exact nature of social influence is unknown, and the total utility of the product is comprised of an individual's utility plus a weighted sum of utilities of others in the social network (see Anselin, 2013 and Bhat, 2014 for applications in spatial econometrics). The *third* and *fourth* approaches assume that individuals revise their attribute importance weight and attribute value itself as a function of preferences of others in their interpersonal network (Narayan et al., 2011).

While threshold-based specification is straightforward to incorporate, they mask the propagation of information in the social network and only offer an aggregate behavioral effect. In the absence of such micro-level dynamics, the interplay between WOM and risk preferences is difficult to model. From the econometric perspective, this approach could lead to biased parameter estimates as it ignores the spatial/social-network-level correlations among individuals. The last two approaches are information hungry as they require precise information about attribute level communication within the social network. Therefore, we mainly rely on the second approach to model social network effects and capture aggregate threshold effects by controlling for the adoption rates of AVs within the individual's social network and the city.

Generally, the exact nature of social influence is not discernable in the second approach, especially when the social network effect is directly incorporated in the utility equation of a choice model (Bhat et al., 2015; Sidharthan and Bhat, 2012). Moreover, such specification does not offer any means for information propagation without directly including several information-related indicators in the utility equation. These additional indicators could induce measurement error and increase the number of covariates substantially. On the other hand, applying the second approach through the latent construct of the ICLV structure has two advantages. First, the essential information related to social influence (i.e., attitudes such as risk-aversion and positive WOM dissemination) can be encapsulated in a low-dimensional vector of latent variables. This specification is also much less vulnerable to measurement errors. Second, the analyst can incorporate the social network effect on latent variables to enable the exchange of information between consumers, a critical trait of a consumer behavior model that makes it useful for forecasting the adoption of novel products (Bhat et al., 2016b). Thus, the ICLV model helps open the black box by putting a structure on the information dissemination (see Bhat et al., 2015, 2016b for ICLV applications). Such behavioral insights derived from ICLV make it an attractive alternative to model social influence. In this study, the information spread by an individual is characterized by the WOM latent variable, which is a function of the WOM of individuals in the social network (see mathematical details in Section 3.3). WOM is measured using responses to indicator questions presented in the WOM experiments (see Figure 2and Figure 3).

## 3.2 Risk Preference

Risk can be broadly classified into seven types – financial, performance, physical, time, social, psychological, and network externality (see Dholakia, 2001; Hirunyawipada and Paswan, 2006 for a detailed review). By measuring the *risk aversion* latent variable through the responses to four risk-related statements summarized in Figure 6, ICLV accounts for *psychological* risks associated with AV adoption. By directly incorporating AV adoption rates at the social network and city level in the utility equation of ICLV, we implicitly model the perceived *time* risk. The information provided in WOM experiments regarding the lack of liability in AV crashes accounts for *performance* risk, and dependence of an individual's WOM on the WOM of social network captures *social* risk. Thus, the proposed interdependent

ICLV model could capture *performance* and *social* risks with the cross-loading of WOM on the risk aversion latent variable.

### 3.3 Interdependent Integrated Choice and Latent Variable Model (ICLV)

Figure 7 details the three components of the interdependent ICLV model – structural equation model, measurement equation model, and discrete choice model. The figure highlights how the WOM of decision-makers is affected by the information obtained from external sources and their interpersonal network. The combined information (represented by WOM) is considered to influence the risk-aversion behavior of decision-makers. Finally, information and risk-aversion attitude, along with product attributes, determine the AV adoption behavior.

We provide a general formulation of all three components of the model and discuss them in the context of the empirical study. Subsequently, we write the joint likelihood function and discuss the estimation details. The methodology is based on the ICLV model of Bhat et al. (2016b), but we make two extensions to the existing model. First, we capture the moderation effect between attitudes in the structural equation through cross-loading of latent variables. Second, the discrete choice component of the ICLV model is adjusted to account for panel effects. We write our own code in GAUSS, a matrix programming language, to estimate the interdependent ICLV model.

*3.3.1 Latent Variable Structural Equation Model*

Let $l$ and $q$ be the indexes for latent variables $l = (1,2,...,L)$ and individuals $q = (1,2,...,Q)$. In the empirical study, we consider WOM ($l=1$) and risk aversion ($l=2$) as two latent variables (i.e., $L=2$). The latent variable $\left(z_{ql}^*\right)$ is written as a linear function of covariates:

$$z_{ql}^* = \alpha_l' s_{ql} + \eta_{ql} \tag{1}$$

Eq. (1) assumes that the individual's latent attitude is independent of other individuals. To incorporate the effect of interpersonal network, Eq. (1) is modified as follows:

$$z_{ql}^* = \alpha_l' s_{ql} + \eta_{ql} + \delta_l \sum_{q'=1}^{Q} w_{qq'} z_{q'l}^* \tag{2}$$

where $s_{ql}$ is a $(F \times 1)$ vector of observed covariates, $\alpha_l$ is the corresponding vector of coefficients, $\eta_{ql}$ is a normally distributed error term, $\delta_l$ $(0 < \delta_l < 1)$ is the autoregressive parameter which captures the interdependence effect across individuals in the interpersonal network, and $w_{qq'}$ is a weight matrix with $w_{qq} = 0$ and $\sum_{q' \neq q}^{Q} w_{qq'} = 1 \ \forall q$. Essentially, Eq. (2) is a spatial auto-correlation regression (Anselin, 2013). In this study, $s_{ql}$ consists of socio-economic characteristics and accident history variables. The variables presented in the WOM experiments are only included in equation corresponding to WOM (i.e., $s_{q1}$). We estimate $\delta_1$ and set $\delta_2$ to zero because social network effects in risk aversion are transmitted through cross-loading of WOM. We define the following notations to write Eq. (2) in matrix form for all $Q$ individuals:

$$\mathbf{z}_q^* = \left(z_{q1}^*, z_{q2}^*, \ldots, z_{qL}^*\right)' \qquad [L \times 1 \text{ vector}],$$

$$\mathbf{z}^* = \left(\left(\mathbf{z}_1^*\right)', \left(\mathbf{z}_2^*\right)', \ldots, \left(\mathbf{z}_Q^*\right)'\right)' \qquad [QL \times 1 \text{ vector}],$$

$$\tilde{\mathbf{s}}_q = \begin{pmatrix} \mathbf{s}_{q1}' & & 0 \\ & \ddots & \\ 0 & & \mathbf{s}_{qL}' \end{pmatrix} \qquad [L \times LF \text{ matrix}],$$

$$\tilde{\mathbf{s}} = \left(\tilde{\mathbf{s}}_1', \tilde{\mathbf{s}}_2', \ldots, \tilde{\mathbf{s}}_Q'\right)' \qquad [QL \times LF \text{ matrix}],$$

$$\boldsymbol{\alpha} = \left(\boldsymbol{\alpha}_1', \boldsymbol{\alpha}_2', \ldots, \boldsymbol{\alpha}_L'\right)' \qquad [LF \times 1 \text{ vector}],$$

$$\boldsymbol{\eta}_q = \left(\eta_{q1}, \eta_{q2}, \ldots, \eta_{qL}\right)' \qquad [L \times 1 \text{ vector}],$$

$$\boldsymbol{\eta} = \left(\boldsymbol{\eta}_1', \boldsymbol{\eta}_2', \ldots, \boldsymbol{\eta}_Q'\right)' \qquad [QL \times 1 \text{ vector}],$$

$$\boldsymbol{\delta} = \left(\delta_1, \delta_2, \ldots, \delta_L\right)' \qquad [L \times 1 \text{ vector}],$$

$$\tilde{\boldsymbol{\delta}} = \mathbf{1}_Q \otimes \boldsymbol{\delta} \qquad [QL \times 1 \text{ vector}],$$

where "$\otimes$" represents the Kronecker product, $\mathbf{IDEN}_L$ is an identity matrix of size $(L \times L)$, and $\mathbf{1}_Q$ is a vector of size $(Q \times 1)$ with all its elements equal to 1. To allow for correlation among the latent variables of an individual, $\boldsymbol{\eta}_q$ is assumed to follow a multivariate normal (MVN) distribution $\boldsymbol{\eta}_q \sim \text{MVN}_L[\mathbf{0}_L, \boldsymbol{\Gamma}]$, where $\mathbf{0}_L$ is an $(L \times 1)$ column vector of zeros, and $\boldsymbol{\Gamma}$ is the correlation matrix of size $(L \times L)$. Note that since variance is normalized to 1 for identification, correlation matrix is the same as covariance matrix. Considering $L = 2$ in this study only $\Gamma_{12}$ is estimable.

We assume $\boldsymbol{\eta}_q$ to be independent across individuals $\left(\text{i.e., Cov}(\boldsymbol{\eta}_q, \boldsymbol{\eta}_{q'}) = 0, \forall q \neq q'\right)$. Thus, Eq. (2) can be written in matrix form for all $Q$ individuals as follows:

$$\mathbf{z}^* = \mathbf{S}\tilde{\mathbf{s}}\boldsymbol{\alpha} + \mathbf{S}\boldsymbol{\eta} \qquad (3)$$

where $\mathbf{S} = \left[\mathbf{IDEN}_{QL} - \tilde{\boldsymbol{\delta}} \cdot * (\mathbf{W} \otimes \mathbf{IDEN}_L)\right]^{-1}$ is a matrix of size $(QL \times QL)$, ".*" represents the product of each element of a vector with the corresponding row of a matrix, $\mathbf{IDEN}_{QL}$ is an identity matrix of size $(QL \times QL)$, and $\mathbf{W}$ is a $(Q \times Q)$ row normalized weight matrix. We construct spatial weight matrices with five and ten social ties using Gower similarity index and Euclidian distance to test sensitivity of results (see Section 3.3.6 for details). To be exact, in the case of five ties, the individual is assumed to be spatially correlated with five nearest neighbors where proximity is measured in terms of the Gower similarity index or spatial distance.

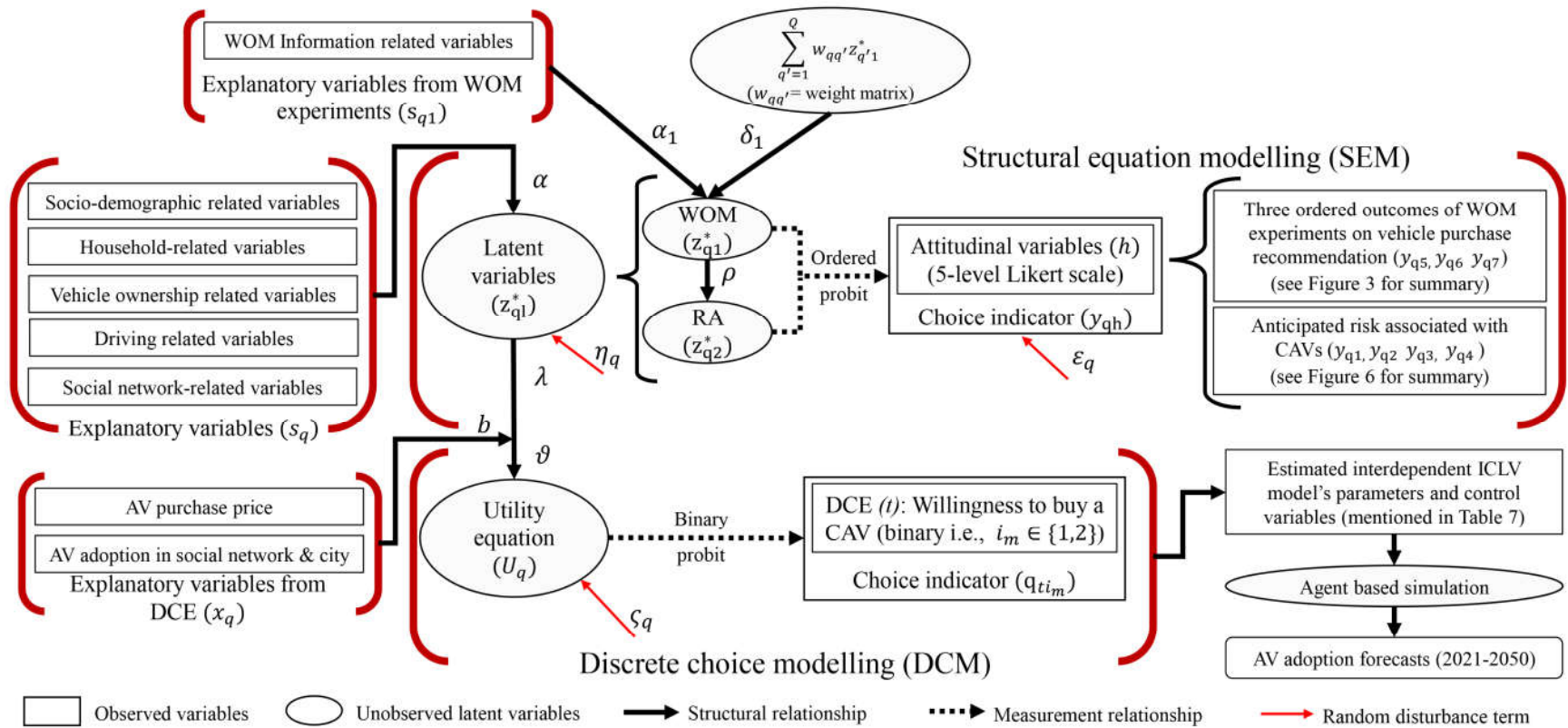

**Figure 7: Modeling framework of the Interdependent ICLV model**

While the attitude of individuals is affected by their interpersonal/spatial network, there may also be a moderation effect across different types of attitudes. For example, if an individual has a strong sense of duty towards the environment (pro-environment), he/she may also exhibit a strong attitude towards trying new environmental-friendly products even if they are novel in the market (risk-taking behavior). Similarly, in the context of this study, WOM can alleviate or augment the risk aversion. Therefore, we extend Eq. (3) to accommodate moderation effect as follows:

We define a matrix $\boldsymbol{\kappa}$ of size $(L \times L)$ with zeros in all cells. If the analyst wishes to load latent variable $l''$ on $l'$, a value of "1" is inserted in the cell $(l', l'')$ of the matrix $\boldsymbol{\kappa}$. For two latent variables in the current study where WOM ($l=1$) is loaded on the risk aversion ($l=2$), the matrix $\boldsymbol{\kappa}$ can be written as follows: $\boldsymbol{\kappa} = \begin{bmatrix} 0 & 0 \\ 1 & 0 \end{bmatrix}$. Next, we define a matrix $\mathbf{R}$ of size $(L \times L)$ with zeros in all cells and place the corresponding cross-loading parameter $\left[\boldsymbol{\rho} = \left(\rho_1, \rho_2, ..., \rho_{(L-1)^2}\right)\right]$ in the cells of matrix $\mathbf{R}$ according to the configuration of matrix $\boldsymbol{\kappa}$. The matrix $\mathbf{R}$ for the case study has the following configuration: $\mathbf{R} = \begin{bmatrix} 0 & 0 \\ \rho_1 & 0 \end{bmatrix}$. With this information, we extend Eq. (3) to account for moderation effects along with spatial effects:

$$z^* = \mathbf{D}[\mathbf{S}\tilde{s}\alpha + \mathbf{S}\eta] \tag{4}$$

where $\mathbf{D} = \mathbf{IDEN}_Q \otimes [\mathbf{IDEN}_L - \mathbf{R}]^{-1}$. Thus, $z^* \sim \text{MVN}_{QL}(\tilde{\boldsymbol{\theta}}, \tilde{\boldsymbol{\Xi}})$, where mean is $\tilde{\boldsymbol{\theta}} = \mathbf{D}\mathbf{S}\tilde{s}\alpha$ and correlation matrix is $\tilde{\boldsymbol{\Xi}} = \mathbf{D}\left(\mathbf{S}\left[\mathbf{IDEN}_Q \otimes \Gamma\right]\mathbf{S}'\right)\mathbf{D}'$.

### 3.3.2 Latent Variable Measurement Equation Model

Since all indicators to capture the underlying attitudes are measured on Likert scale, we only present the measurement equation system corresponding to ordinal variables. Readers are referred to Bhat et al. (2016a) for a comprehensive measurement equation system with a combination of continuous, ordinal, count, and nominal outcomes.

Let $h$ be the index for ordinal variables $(h = 1, 2, ..., H)$ and $J$ be the number of categories for ordinal outcomes. In this study, $H = 7$ (3 for WOM [see Figure 3] and 4 for risk aversion [see Figure 6]) and $J = 5$. Let $y_{qh}^*$ be the latent variable which leads to the observed outcome $y_{qh}$ for individual $q$ and ordinal variable $h$. Following the usual ordered response formulation, we can write the link function:

$$y_{qh}^* = \boldsymbol{\gamma}_h' \mathbf{x}_{qh}^* + \boldsymbol{d}_h' \mathbf{z}_q^* + \varepsilon_{qh}, \qquad \psi_{h, y_{qh}-1} < y_{qh}^* < \psi_{h, y_{qh}}, \tag{5}$$

where $\mathbf{x}_{qh}^*$ is a $(K \times 1)$ vector of observed covariates (including the constant), $\boldsymbol{\gamma}_h$ is the corresponding vector of coefficients, $\boldsymbol{d}_h$ is a $(L \times 1)$ vector of latent variable loadings on the ordinal variable $h$, $\varepsilon_{qh}$ is a normally distributed random error term, and $\psi_{h, y_{qh}}$ is the threshold. For each ordinal variable, the thresholds should be in ascending order and should cover the real line, i.e. $\psi_{h,0} < \psi_{h,1} < \psi_{h,2} ... < \psi_{h,J-1} < \psi_{h,J}$, where $\psi_{h,0} = -\infty$ and $\psi_{h,J} = \infty$. To set the origin of the ordinal variable, one can either set the second threshold or the intercept to zero. Here, we choose to do the former: $\psi_{h,1} = 0$. In the empirical study, $x^*_{qh}$ only has a constant $(K = 1)$. Next, we define the following notations to write Eq. (5) in a matrix form.

$$y^*_q = \left(y^*_{q1}, y^*_{q2}, ..., y^*_{qH}\right)' \qquad [H \times 1 \text{ vector}],$$

$$y_q = \left(y_{q1}, y_{q2}, ..., y_{qH}\right)' \qquad [H \times 1 \text{ vector}],$$

$$\tilde{\gamma} = \left(\gamma_1, \gamma_2, ..., \gamma_H\right)' \qquad [H \times K \text{ matrix}],$$

$$x^*_q = \left(x^*_{q1}, x^*_{q2}, ..., x^*_{qH}\right)' \qquad [H \times K \text{ matrix}],$$

$$d = \left(d_1, d_2, ..., d_H\right)' \qquad [H \times L \text{ matrix}],$$

$$\varepsilon_q = \left(\varepsilon_{q1}, \varepsilon_{q2}, ..., \varepsilon_{qH}\right)' \qquad [H \times 1 \text{ vector}],$$

$$\psi_h = \left(\psi_{h,0}, \psi_{h,1}, ..., \psi_{h,J}\right)' \qquad [J \times 1 \text{ vector}],$$

$$\psi = \left(\psi'_1, \psi'_2, ..., \psi'_H\right)' \qquad [HJ \times 1 \text{ vector}],$$

$$\psi_{q,low} = \left(\psi_{1,y_{q1}-1}, \psi_{2,y_{q2}-1}, ..., \psi_{H,y_{qH}-1}\right)' \qquad [H \times 1 \text{ vector}],$$

$$\psi_{q,up} = \left(\psi_{1,y_{q1}}, \psi_{2,y_{q2}}, ..., \psi_{H,y_{qH}}\right)' \qquad [H \times 1 \text{ vector}],$$

Eq. (5) can be written in matrix form for individual $q$ as follows:

$$y^*_q = \text{sumc}\left(\tilde{\gamma} .* x^*_q\right) + dz^*_q + \varepsilon_q, \quad \psi_{q,low} < y^*_q < \psi_{q,up} \qquad (6)$$

where "sumc" implies sum across columns of the matrix and $(.*)$ denotes the dot product of matrices. To reduce the model complexity, $\varepsilon_q$ is assumed to follow a standard MVN distribution: $\varepsilon_q \sim \text{MVN}_{H \times H}(\mathbf{0}_H, \mathbf{IDEN}_H)$ because the latent variables ($z^*_q$) loadings naturally generate the correlation across ordinal variables.

### 3.3.3 Discrete Choice Model

Let $t$ be the index for choice occasion $(t = 1, 2, ..., T)$, and $i$ be the index for alternative $(i = 1, 2, ..., I)$. In the case study, $T = 3$ and $I = 2$. The indirect utility of individual $q$ due to choosing alternative $i$ during choice occasion $t$ is:

$$U_{qti} = b'x_{qti} + \lambda'_i z^*_q + \vartheta'(\mathbf{A}_{qti} z^*_q) + \varsigma_{qti} \qquad (7)$$

where $x_{qti}$ is an $(M \times 1)$ vector of alternative-specific attributes (including the constant), $b$ is a vector of corresponding marginal utilities, $\lambda_i$ is a $(L \times 1)$ vector of coefficients of latent variables for alternative $i$, and $\varsigma_{qit}$ is a normally-distributed error term. In the empirical study, $x_{qti}$ consists of alternative-specific constant, purchase price, percentage adoption in the social network, and percentage adoption in the city (i.e., $M=4$). Note that $\lambda_i$ for one alternative is normalized to zero for identification. $\mathbf{A}_{qti}$ is a $(LM \times L)$ matrix of altenative-specific attributes $(x_{qti})$ that interact with latent variables, and $\vartheta$ is the corresponding $(LM \times 1)$ column vector of marginal utilities. We define the following notations to convert Eq. (7) into matrix form:

$$U_{qt} = (U_{qt1}, U_{qt2}, ..., U_{qtI})' \quad [I \times 1 \text{ vector}],$$

$$U_q = (U'_{q1}, U'_{q2}, ..., U'_{qT})' \quad [TI \times 1 \text{ vector}],$$

$$x_{qt} = (x_{qt1}, x_{qt2}, ..., x_{qtI}) \quad [M \times I \text{ matrix}],$$

$$x_q = (x_{q1}, x_{q2}, ..., x_{qT})' \quad [TI \times M \text{ matrix}],$$

$$\tilde{b} = (b, b, ..., b)' \quad [TI \times M \text{ matrix}],$$

$$\lambda = (\lambda_1, \lambda_2 ..., \lambda_I) \quad [L \times I \text{ vector}],$$

$$\tilde{\lambda} = (\lambda, \lambda, ..., \lambda)' \quad [TI \times L \text{ vector}],$$

$$\varpi_{qti} = (\vartheta' \mathbf{A}_{qti}) \quad [1 \times L \text{ vector}],$$

$$\varpi_{qt} = (\varpi'_{qt1}, \varpi'_{qt2}, ..., \varpi'_{qtI}) \quad [L \times I \text{ matrix}],$$

$$\varpi_q = (\varpi_{q1}, \varpi_{q2}, ..., \varpi_{qT})' \quad [TI \times L \text{ matrix}],$$

$$\varsigma_{qt} = (\varsigma_{qt1}, \varsigma_{qt2}, ..., \varsigma_{qtI}) \quad [1 \times I \text{ vector}],$$

$$\varsigma_q = (\varsigma_{q1}, \varsigma_{q2}, ..., \varsigma_{qT})' \quad [TI \times 1 \text{ vector}],$$

Let $\mathbf{\Lambda}$ be the covariance matrix of $\varsigma_{qt}$, which we assume to be independent across choice occasions to reduce the model complexity because latent variable ($z_q^*$) loadings generate the correlations across choice occasions. Thus, Eq. (7) for individual $q$ can be written as follow:

$$U_q = \text{sumc}(\tilde{b}.*x_q) + (\tilde{\lambda} + \varpi_q)z_q^* + \varsigma_q = \text{sumc}(\tilde{b}.*x_q) + \tilde{\tau}_q z_q^* + \varsigma_q, \tag{8}$$

where "sumc" implies sum across columns of the matrix, $(.*)$ denotes the dot product of matrices, and $\varsigma_q \sim MVN_{TI \times TI}(\mathbf{0}, \mathbf{IDEN}_T \otimes \mathbf{\Lambda})$.

### 3.3.4 Joint Likelihood

Note that obtaining the marginal distribution of $U_q$ and $y_q^*$ is not straighford due to spatial correlation in the latent variable ($z_q^*$) across individuals. Therefore, we work with the joint distribution of all individuals to write the model likelihood:

$$y^*U \sim MVN_{[Q(H+TI)]\times[Q(H+TI)]}\left(\mathbf{B} = \mathbf{\mu}_1 + \mathbf{\mu}_2\tilde{\mathbf{\theta}}, \mathbf{\Omega} = \mathbf{\mu}_2\tilde{\mathbf{\Xi}}\mathbf{\mu}_2' + \mathbf{IDEN}_Q \otimes \mathbf{\Sigma}\right), \quad (9)$$

where

$$\mathbf{\mu}_{1q} = \left(\left[\text{sumc}(\tilde{\mathbf{\gamma}} .* \mathbf{x}_q^*)\right]', \left[\text{sumc}(\tilde{\mathbf{b}} .* \mathbf{x}_q)\right]'\right)' \qquad [(H+TI)\times 1 \text{ vector}],$$

$$\mathbf{\mu}_1 = \left(\mathbf{\mu}_{11}', \cdots, \mathbf{\mu}_{1Q}'\right)' \qquad [Q(H+TI)\times 1 \text{ vector}],$$

$$\mathbf{\mu}_{2q} = \left(\mathbf{d}', \tilde{\mathbf{\tau}}_q'\right)' \qquad [(H+TI)\times L \text{ matrix}],$$

$$\mathbf{\mu}_2 = \begin{pmatrix} \mathbf{\mu}_{21} & & 0 \\ & \ddots & \\ 0 & & \mathbf{\mu}_{2Q} \end{pmatrix} \qquad [Q(H+TI)\times QL \text{ matrix}],$$

$$\mathbf{\Sigma} = \begin{pmatrix} \mathbf{IDEN}_H & \mathbf{0}_{H\times TI} \\ \mathbf{0}_{TI\times H} & \mathbf{IDEN}_T \otimes \mathbf{\Lambda} \end{pmatrix} \qquad [(H+TI)\times(H+TI) \text{ matrix}],$$

Since, only the difference in utility matters, we work with utility differences in the discrete choice part. We specifically subtract the utility of the chosen alternative from utilities of all non-chosen alternatives. Moreover, top left element of the differenced error covariance matrix ($\tilde{\mathbf{\Lambda}}$) is fixed to 1 to set the utility scale for identifiability (Keane, 1992). Thus, for $I$ alternatives case, only $\{I*(I-1)*0.5\}-1$ covariance elements are identifiable. Further, since all the differenced error covariance matrices must originate from the same undifferenced error covariance matrix ($\mathbf{\Lambda}$), we specify matrix $\mathbf{\Lambda}$ as follows (Sidharthan and Bhat, 2012): $\mathbf{\Lambda} = \begin{bmatrix} 0 & 0 \\ 0 & \tilde{\mathbf{\Lambda}} \end{bmatrix}$. We also had to transform Eq. (9) in the utility-difference space using a matrix $\mathbf{M}$ of size $[Q(H+T(I-1))\times Q(H+TI)]$ as shown in Eq. (10). The procedure to compute $\mathbf{M}$ is provided in Algorithm A.1 in Appendix 1. The resulting joint distribution in the utility differenced space is:

$$\bar{y}^*\bar{U} \sim MVN_{[Q(H+T(I-1))]\times[Q(H+T(I-1))]}\left(\bar{\mathbf{B}} = \mathbf{M}B, \bar{\mathbf{\Omega}} = \mathbf{M}\mathbf{\Omega}\mathbf{M}'\right) \quad (10)$$

*3.3.5 Estimation*

To evaluate the joint likelihood presented in Eq. (10), we need to compute a $Q(H+T(I-1))$ dimensional integral involved in the multivariate normal cumulative distribution function (MVNCDF). Despite several advancements in quasi-Monte Carlo (Bhat, 2003), quadrature (Bansal et al., 2021), analytical methods (Bhat, 2018) to evaluate high-dimensional integrals, computing such a high-dimensional integral at the desired accuracy remains infeasible. Therefore, we make use of the composite marginal likelihood (CML) approach, which define a surrogate likelihood function by taking the product of low dimensional MVNCDF. The surrogate function for Eq. (10) is as follows:

$$L_{CML}(\Theta) = \left[\prod_{q=1}^{Q}\prod_{h=1}^{H}\prod_{q'=1}^{Q}\prod_{h'=1}^{H}\Pr\left[y_{qh}, y_{q'h'}\right]\right] \times \left[\prod_{q=1}^{Q}\prod_{q'=1}^{Q}\prod_{h=1}^{H}\prod_{t=1}^{T}\Pr\left[y_{qh}, q'_{ti_m}\right]\right]$$

$$\left[\prod_{q=1}^{Q}\prod_{t=1}^{T}\prod_{q'=q+1}^{Q}\prod_{t'=t+1}^{T}\Pr\left[q_{ti_m}, q'_{t'i_m}\right]\right] \tag{11}$$

where $\Theta = [Vech(\alpha), Vech(\Gamma), \delta, \rho, Vech(\tilde{\gamma}), Vech(d), \psi, b, Vech(\lambda), \vartheta, Vech(\Lambda)]$,

"$Vech(.)$" vectorizes all the elements of the matrix, $y_{qh}$ indicates the observed scale for the ordinal variable $h$ by the individual $q$, and $q_{ti_m}$ indicates the chosen alternative $i_m$ at the choice occasion $t$ by the individual $q$. In Eq. (11), for each pair of individuals $q$ and $q'$, the first, second, and third terms correspond to the pairing of ordinal variables, pairing of ordinal variables with nominal variables, and pairing of nominal variables, respectively. Further details about computing surrogate function are provided in Appendix 1.

The analyst needs to calculate the robust asymptotic covariance matrix, a function of Jacobian and Hessian matrices, to obtain standard errors of the model parameters (Godambe, 1960). The calculation of Jacobian and Hessian matrices for choice models with interdependencies is different from the traditional choice models. Readers are referred to Sidharthan and Bhat (2012) for a discussion on the calculation of the sandwich covariance matrix.

### 3.3.6 Spatial Weight Matrix Computation

All the data required to estimate the interdependent ICLV model can be obtained from a stated preference survey, except the weight matrix used in Eq. (2). Essentially, the weight matrix determines the un-moderated weights assigned by individuals to others in their interpersonal network. The weight matrix can be constructed in two ways. The first method relies on the concept that "everything is related to everything else, but near things are more related than distant things" (Tobler, 1970). This concept has been extensively used in spatial econometrics to construct the weight matrix as a function of spatial distance, such as the inverse of distance (Anselin, 2010). On the other hand, the second method uses the concept of homophilic – similarity in terms of socio-demographic attributes, proximity, attitudes, and other behavioral characteristics. This concept is often used in social sciences to model information propagation and online interactions (Bhattacharya and Sarkar, 2021; David-Barrett, 2020).

In the current study, we construct weight matrices using both methods – i) geographical distance (inverse of distance) and ii) homophilic based weight matrices. For the geographical distance-based weight matrix, the strength of a tie between two individuals is negatively proportional to the Euclidian distance between their residential locations. We use the R package "geosphere" to calculate the shortest pair-wise distance between five-digit zip codes of home locations (Hijmans et al., 2017). We compute the homophilic-based index using the Gower dissimilarity index because it can handle both continuous and categorical data on socio-demographic characteristics (Gower, 1971). The Gower dissimilarity index $w_{qq'}$ is given by:

$$w_{qq'} = \frac{\sum_{k=1}^{K} \omega_k s_{qq'k}}{\sum_{k=1}^{K} \omega_k} \tag{12}$$

where $s_{qq'k}$ is the contribution of the variable $k$ in the similarity between individuals $q$ and $q'$ $(q \neq q'; q, q' \in Q)$ and $\omega_k$ is the corresponding weight. For the continuous and categorical variables, contributions are calculated using normalized Manhattan and Dice distances, respectively. A value of $w_{qq'}$ close to zero indicates a strong tie. We consider socio-demographic characteristics like age, gender, ethnicity, educational attainment, and household income to generate the Gower-distance-based weight matrix. The contribution weight $\omega_k$ for each variable is assumed to be the same.

Further, a closer look at Eq. (11) indicates that the likelihood computation requires the joint probability calculation for all pairs of individuals. Assuming that each individual is connected with every other individual is unrealistic and computationally prohibitive. To make the likelihood estimation tractable, one can either follow Tobler's law (i.e., weight beyond a certain distance is zero) or put constraints on the number of effective social ties in an individual's interpersonal network. In this study, we use the latter and consider two configurations with five and ten social ties. Thus, we estimate the interdependent ICLV model with four specifications of the weight matrix and ties -- spatial distance with five ties, spatial distance with ten ties, Gower distance with five ties, and Gower distance with ten ties.

## 4. Results of Consumer Behavior Model

### 4.1 Statistical Model fit assessment

As we estimate the ICLV model using the CML approach, we adopt composite likelihood information criterion (CLIC) to compare non-nested models, i.e. models with the same number of ties but different weight matrix configurations (Varin and Vidoni, 2005). CLIC is computed using the following expression: $\left[ \log L_{CML}(\Theta) - \text{tr}\left( J(\Theta) H(\Theta)^{-1} \right) \right]$, where $\left[ \log L_{CML}(\Theta) \right]$ is the composite log-likelihood and $\text{tr}\left[ J(\Theta) H(\Theta)^{-1} \right]$ is the penalty term (trace of the product of Jacobian and Hessian inverse). The model with higher CLIC is preferred. Table 3 provides the CLIC statistics and other relevant statistics for all the four weight matrix specifications.

Table 3: Summary of model fit statistics

| Specification | | Composite Log-likelihood | Composite likelihood information criterion (CLIC) | Choice model Composite log-likelihood | Average choice model Composite log-likelihood |
|---|---|---|---|---|---|
| Distance measure | Number of ties | | | | |
| Gower | 5 | -2,732,802.7 | -2,744,537.6 | -1,32,590.1 | -1.185 |
| Spatial | 5 | -2,738,387.5 | -2,744,228.7 | -132,771.6 | -1.187 |
| Gower | 10 | -5,455,102.7 | -5,467,804.5 | -264,796.9 | -1.185 |
| Spatial | 10 | -5,458,220.9 | -5,529,365.6 | -264,199.3 | -1.182 |

The weight matrix specification based on spatial distance has a better model fit for the five-tie case, but the specification based on Gower distance outperforms the one with spatial distance in the ten-tie scenario. However, the comparison of choice model composite log-likelihood values of specifications with the same ties but different distance measures suggests that the difference is not substantial – 182 and 598 point difference in five and ten ties scenarios, respectively. We also compute the average choice model composite log-likelihood value by dividing total CML value with the number of effective pairs $\left\{ \sum_{q=1}^{Q} \left[ \sum_{q'=1}^{\#\text{of ties}} \sum_{t=1}^{T} \sum_{t'=1}^{T} \right] \right\}$ to compare all specification. The results show that the model fit is only marginally sensitive to the configuration of the spatial weight matrix.

### 4.1.1 Structural Equation Model Results

Table 4 provides the parameter estimates of the structural equation for latent variables. Since the direction of effects does not change much across weight matrix specifications, we present and discuss results for the weight matrix based on Gower distance with five ties. The results of other three weight matrix configurations and the non-spatial model are provided in online supplementary material.

Several demographic characteristics, type and source of information, and accident history have a statistically significant effect on WOM dissemination. For instance, bachelor's degree holders are more likely to spread positive information about AVs than those with lower education levels, ceteris paribus. The positive effect of education could be attributed to the fact that highly educated individuals can process more information and thereby be more certain about the long-term benefits of AVs, such as lower accident rates and emissions (Golbabaei et al., 2020; Haboucha et al., 2017; Han et al., 2011; Jansson et al., 2011; Jerit et al., 2006; Knight et al., 2010; Liljamo et al., 2018). Individuals living in high-income households tend to be more active in disseminating positive information than their low-income counterparts (Bansal et al., 2016). The effect of income on spreading positive WOM could be related to hedonic experience (Paridon et al., 2006). In terms of household configurations, three demographic variables have a positive effect on WOM dissemination. The number of workers in the household has a positive impact on WOM communication. One possible reason for such positive WOM communication could be the self-relocation and parking capability of the AVs, a trait beneficial for large working households (Baron et al., 2021). The number of children in the household is also linked to the positive dissemination of WOM communication, possibly due to the enhanced safety features offered by AVs (Sinha et al., 2020). The results also suggest that male decision-makers are more likely to spread positive WOM compared to their female counterparts (Kim et al., 2019; Liljamo et al., 2018; Zoellick et al., 2019). A higher propensity of men towards AVs could be attributed to status symbols. Owning an AV can convey a sense of symbol or power manifested through a willingness to pay a premium price for new innovative car technology (Wadud and Chintakayala, 2021).

**Table 4: Structural equation model parameter estimates for interdependent ICLV model with weight matrix based on Gower distance with five ties**

| Explanatory Variables | | Coefficient (t-stat) | |
|---|---|---|---|
| | | Word of Mouth (WOM) | Risk Aversion |
| Education Status | Some college degree or below | -0.112 (-6.61) | --- |
| | Professional degree (MD, JD, etc.) | -0.190 (-7.93) | --- |
| | College Graduate | --- | 0.245 (7.51) |
| | MS, PhD or Doctoral degree | --- | 0.245 (7.51) |
| Household Income Base: >75K | <=35K | -0.283 (-4.29) | --- |
| | 36K - 75K | -0.151 (-15.13) | --- |
| Household Configuration | Number of Workers | 0.273 (20.01) | 0.067 (3.28) |
| | Number of Children | 0.279 (18.92) | 0.054 (2.02) |
| | Respondent Male (Base: Female) | 0.061 (3.55) | -0.197 (-6.86) |
| Vehicle Ownership | Number of Vehicles | -0.127 (-17.19) | --- |
| Information Variables | Number of crashes in AV | -0.044 (-13.99) | --- |
| | Reduction in travel time in AV | --- | --- |
| | Reduction in $CO_2$ emission in AV | --- | --- |
| | Unclear liability in x% of AV-involved crashes. | -0.624 (-14.06) | --- |
| Information Source Base: Car Dealer | Friend | 0.009 (1.88) | --- |
| | Colleague at work | 0.031 (3.27) | --- |
| | Media | 0.092 (9.03) | --- |
| Past One year accident involvement. Number of accidents where… | Vehicle incurred minor damages. | -0.060 (-5.29) | --- |
| | Vehicle incurred major damages. | 0.039 (4.11) | --- |
| | I suffered from minor injuries. | -0.136 (-11.66) | --- |
| | I suffered from severe injuries. | 0.084 (7.86) | --- |
| Cross Loading $(\rho)$ | WOM | | -1.07 (-2.25) |
| Spatial Parameter $(\delta)$ | | 0.244 (2.01) | 0 (fixed) |
| Correlation between WOM and Risk Aversion $(\Gamma_{12})$ | | 0.394 (4.18) | |

---: indicates that the parameter was not significant at a significance level of 0.2 and hence removed

Vehicle ownership (i.e., the number of cars) is negatively related to WOM dissemination (Liljamo et al., 2018). Since vehicle ownership is generally a proxy for driving propensity (Kaneko and Kagawa, 2021), households with higher vehicle ownership may not want to relinquish the pleasure derived from manual driving. Two of the four information variables are found to be statistically significant in explaining the WOM, namely the number of AV-involved crashes and the proportion of such crashes with the lack of liability. Both covariates have an intuitive sign. We include informational variables in the structural equation part of the model so that their propagation effect through the social network can be explicitly captured. This structural relationship also leads to an elegant top-down propagation of the effect of information from latent variables to observed indicators and choices.

Among information sources, information received from friends, colleagues, and media has a positive influence on WOM compared to that of the car dealer. This result might be a consequence of relatively lesser trust in the information provided by dealers. Finally, the severity of injury/damage in accidents has a stimulating effect on WOM dissemination. In the event of minor injury/damage to the individual/vehicle, the effect is negative. However, the effect turns out to be positive in the event of major injury/damage. The results indicate that people with the worst experience of manual driving are more likely to realize the benefits of automated driving from the safety perspective (Menon et al., 2019).

We now discuss the parameter estimates of the structural equation for the second latent variable – risk aversion. Bachelor's degree holders exhibit higher risk aversion compared to their counterparts with lower education (Jung, 2015), perhaps because they are more aware of the risks associated with the adoption of nascent AV technology. Moreover, females with a higher number of workers and children in the household are likely to be highly risk-averse compared to their counterparts. The cross-loading parameter $(\rho)$ (i.e., loading of WOM on risk aversion) is negative and statistically significant, suggesting that the increase in positive WOM reduces risk aversion. The result is aligned with intuition. Note that the actual association of demographic characteristics and risk-averse behavior should be derived after considering the indirect effect of these variables on risk aversion through WOM.

The social network effect is captured through an autoregressive parameter $(\delta)$ in the structural equation of WOM. The autoregressive parameter is significant and has a value of 0.26. This parameter estimate can be interpreted as the weight given by individuals to the information received from their social network. As discussed earlier, we do not estimate the autoregressive parameter for risk aversion because cross-loading of WOM $(\rho)$ implicitly capture social network effect. Finally, the error correlation $(\Gamma_{12})$ between WOM and risk aversion is 0.27, suggesting that the aggregate effect of unobserved variables on both latent variables is in the same direction.

### 4.1.2 *Measurement Equation Model Results*

The parameter estimates for the measurement equation of latent variables are provided in Table 5. Two out of three measurement indicators of WOM and all four measurement indicators of risk aversion show association. All the loading parameters have the expected signs. For example, WOM has positive loading on the statement "I will suggest them to consider buying an AV over a CV because the former is much safer" measured on a five-point Likert scale going from strongly disagree to strongly agree. Other loading parameters can be interpreted similarly. Table 5 also provides the estimates for the intercept and thresholds of ordered indicators. Very small standard errors of thresholds indicate that cut-off values are statistically separated from each other at a significance level of 0.05.

Table 5: Measurement equation model parameter estimates for interdependent ICLV model with weight matrix based on Gower distance with five ties, $\psi_1 = -\infty, \psi_2 = 0, \psi_6 = \infty$

| Statement (Five-point Likert scale with labels strongly disagree to strongly agree) | Coefficient (t-stat) | | Coefficient (standard error) | | | |
|---|---|---|---|---|---|---|
| | Word of Mouth (WOM) | Risk Aversion | Intercept | $\psi_3$ | $\psi_4$ | $\psi_5$ |
| I will suggest them to consider buying an autonomous car over a conventional car because the former is much safer. | 1.160 (17.02) | | 1.554 (12.14) | 0.990 (0.06) | 2.402 (0.13) | 3.857 (0.19) |
| I will suggest them to consider buying a conventional car over an autonomous car because at least one knows who is responsible for a crash in a conventional car. | -0.338 (-8.70) | | 1.958 (9.63) | 0.975 (0.02) | 2.011 (0.03) | 2.890 (0.04) |
| I am worried that I might not get value-for-money in an autonomous car purchase. | | 0.276 (2.24) | 1.524 (14.85) | 0.673 (0.02) | 1.438 (0.02) | 2.438 (0.03) |
| I would take the risk with autonomous car purchase in exchange for an exciting and novel experience. | | -0.574 (-2.22) | 0.807 (23.32) | 0.762 (0.02) | 1.692 (0.04) | 2.583 (0.06) |
| I would feel uncomfortable in switching to autonomous cars. | | 1.055 (2.27) | 2.621 (52.76) | 1.149 (0.07) | 2.073 (0.12) | 3.482 (0.19) |
| I struggle in taking risks with such unconventional decisions. | | 0.616 (2.26) | 1.793 (40.87) | 0.945 (0.03) | 1.872 (0.06) | 2.907 (0.10) |

*4.1.3 Choice Model Results*

Table 6 summarizes the parameter estimates of the utility equation of the binary choice variable – whether to buy an AV or not. Aligned with the intuition, the effects of social and city-level adoption rates on the individual's likelihood to adopt an AV are positive and statistically significant (Bansal et al., 2016; Bansal and Kockelman, 2018; Sharma and Mishra, 2020). The effect of price also exhibits the expected trend, as the likelihood of buying an AV decreases with an increase in the AV price relative to the CV price. We consider a non-linear specification with an estimable power parameter on the ratio of AV price and CV price. However, the price effect turns out to be linear because the power parameter is not statistically different from 1 (at a significance level of 0.2).

The effect of latent variables on the likelihood of buying an AV has an intuitive sign. For instance, individuals with positive WOM dissemination and lower risk aversion have a higher inclination towards buying AVs. Further, we also explore the interaction effects of latent variables and the adoption rates at both social network and city levels. The interaction parameter estimates indicate that the effect of positive WOM dissemination increases with the increase in city-level AV adoption, whereas the negative effect of risk aversion is pacified by the AV adoption in the social network. Finally, the WOM latent variable also has a statistically significant interaction effect with the ratio of AV price to CV price. The positive interaction effect parameter indicates that the negative effect of AV price decreases with the increase in positive WOM dissemination.

Table 6: Choice model parameter estimates for interdependent ICLV model with weight matrix based on Gower distance with five ties (base: will not buy an AV)

|  | Explanatory Variables | Coefficient (t-stat) |
|---|---|---|
| Price and adoption variables | Constant | 0.987 (2.08) |
|  | Percentage adoption in social ties | 1.091 (2.09) |
|  | Percentage adoption in city/community | 0.518 (2.13) |
|  | Ratio of AV to CV price | -2.529 (-3.38) |
| Latent Variable Loading | WOM | 0.596 (2.56) |
|  | Risk Averse | -0.634 (-2.26) |
| Latent Variable Interaction | WOM * Percentage adoption in social ties | --- |
|  | Risk Averse * Percentage adoption in social ties | 1.125 (3.18) |
|  | WOM * Percentage adoption in city/community | 0.395 (4.12) |
|  | Risk Averse * Percentage adoption in city/community | --- |
|  | WOM * Ratio of AV to CV price | 0.672 (2.91) |
|  | Risk Averse * Ratio of AV to CV price | --- |

---: indicates that the parameter was not significant at a significance level of 0.2 and hence removed

While the parameter estimates of the consumer behavior model provide the directional effect of several variables on the likelihood of AV adoption, they cannot be translated into analytical expressions of the forecasted market share of AVs. To this end, we use these parameter estimates and perform an agent-based simulation to understand the market evolution of AV under various scenarios (e.g., change in AV price and reduction in AV-involved crash rates).

## 5. Agent-Based Model: simulation

### 5.1 Synthetic Population Generation

To run an agent-based simulation for the entire household population of Nashville (Davidson County), we expand the survey sample to the household-level synthetic population using an iterative proportional updating (IPU) algorithm. The algorithm adjusts and reallocates weights among households until household- and person-level attributes are both matched with the marginal distributions of attributes in the population (Konduri et al., 2016). We obtain population-level marginal distribution (at household- and person-level) for Nashville from American Community Survey 2013-2017 (Manson et al., 2019). We consider household size, income, and the number of workers, children, and vehicles as the household-level control variables. In addition, age, gender, ethnicity, educational attainment, and disability are used as person-level control variables. To match the spatial distribution of the population, we utilize the five-digit ZIP code of the survey respondent's home location and census tract from the population using their crosswalk (Din and Wilson, 2020). We apply the IPU algorithm through open-source Python software, PopGen (Konduri et al., 2016). The survey sample of 1,495 respondents is thus expanded to a synthetic population of 421,223 households.

## 5.2 Agent-Based Simulation Framework

**Error! Reference source not found.** presents the agent-based simulation (ABM) framework, which takes the estimated consumer behavior model, synthetic population, agent-level social network and the control variables mentioned in Table 7 as inputs. The ABM is run for 30 iterations where the iteration corresponds to one year. The AV price at time $t$ is obtained using the following discounting equation: starting price $\times$ (1 – yearly discount rate)$^{(t-1)}$. The two variables related to AV safety and regulation – the number of AV-involved crashes and the proportion of AV-involved crashes with unknown liability – are generated based on their range in the WOM choice experiment. The magnitude of both variables is iteratively reduced to represent the expectations of experts that AV-involved crashes would reduce with the advancement in the automation technology and stronger laws will be developed to ensure liability of AV-involved crashes. Specifically, we adopt the following quadratic function to find the value of these variable at time $t$ (see Table 1 for upper limits): Upper limit $-\left(\sqrt{\text{Upper limit}}\right)*(t/\text{Curvature})-1$. Moreover, the percentage of the population receiving the information about the above-discussed two information variables through media in year $t$ is determined by the following function: $2\sqrt{t}+2t+1$. This function ensures an increase in media exposure over time. For example, the function implies that 5% people receive information through media in the first year, but the proportion jumps to 15% in the fifth year and 72% in the thirtieth year. Finally, an agent also gets information from *friend* when the AV adoption in her social network exceeds 40%.

    By inserting the above-discussed inputs, spatial weight matrix, agent's social network, and other characteristics of agents in Eq. (4), we calculate the values of both latent variables – WOM and risk aversion – for each agent in a year. While applying Eq. (4), we calculate WOM followed by risk aversion to ensure the loading of WOM on risk aversion is explicitly accounted. By plugging in latent variable values, the purchase price, and city-level and social-network-level AV adoption in Eq. (8), we compute the systematic part of utility. Subsequently, the probability of choosing the AV by each agent in a year is obtained. The probability calculation involves the computation of the cumulative distribution function of normal distribution. The *probability* of adopting an AV is translated into the *decision to buy* an AV if the probability is greater than the random number drawn from a uniform distribution with a range (0.5, 1).

Table 7: Control variables and their description

| Variable | Description |
|---|---|
| Starting price | Starting market price of AVs |
| Yearly discount rate | Discount rate which determines the annual reduction in the AV price |
| Proportion of satisfied agents | Percentage of agents who are satisfied with the purchase of AV |
| Curvature crash | Curvature value used in the function to determine the rate of reduction in AV-involved crashes |
| Curvature legal | Curvature value used in the function to determine the rate of reduction in legal issues related to AV-involved crashes |

Once an agent purchases an AV, the probability calculation for such agents is not performed in future iterations, but their post-purchase AV experience (positive/negative) is updated only once in the subsequent year after the purchase. Since there is no systematic way to capture post-purchase experiences of novel products like AVs, we use a random number generation approach. Specifically, we fix the proportion of agents with post-purchase satisfaction at the beginning of the simulation (see the variable *proportion of satisfied agents* in Table 7). We draw a random number from standard uniform distribution and assign positive experience to an agent if the random number is below the pre-specified proportion of satisfied agents, else negative experience is assigned. Depending upon the type of experience (positive or negative), an agent is assigned with a maximum or a minimum of her social network's WOM value. This WOM updating ensures that the type of post-purchase experience is fixed for an agent, but the WOM value (a proxy for the extent of positive/negative experience) may evolve over years.

Finally, social network- and city-level AV adoption levels are updated at the end of each year. The entire simulation is repeated for 30 iterations (i.e., 2021-2050) for any given scenario using ten different starting values (i.e., random seeds). Since the standard deviation of the forecasted market shares across different starting values is small, we only plot the average of the city-level AV adoption across different starting values. Multiple simulation scenarios are considered by varying control variable values in Table 7.

### 5.3 Agent-Based Simulation Results

We forecast the market share of AVs on a thirty-year horizon (2021-2050) in three scenarios characterized by a combination of i) AV price, ii) the extent of the post-purchase satisfaction of early adopters, iii) reduction in AV-involved crash rates, and iv) clear liability in AV-involved crashes. All three scenarios are simulated for four different social network configurations of social networks, i.e., a combination of the two social distance measures (Euclidian and Gower) and two values of social ties (five and ten). We discuss the ABM results for the social network created based on Gower distance and five ties. The results of the other three social network configurations are provided as online supplementary material.

*5.3.1  Scenario 1: Effect of AV price and post-purchase satisfaction of adopters*

We consider starting price of AVs to be $40,000 and assume it to decrease annually by 1%, 5%, and 10% (Bansal and Kockelman, 2018). Moreover, we forecast AV market share for the proportion of adopters with a positive post-purchase experience being either 30% or 90%. The plots of the forecasted AV adoption are shown in Figure 9(a)**Error! Reference source not found.**. The results indicate that price reduction of the AV technology has a strong effect on AV adoption. Whereas a 5% annual reduction in AV price can help achieve a market share of around 80-85% in the next thirty years, an annual reduction of 1% would lead to an AV market share of only around 15-20% in 2050. The higher post-purchase satisfaction also leads to quicker AV adoption and improves the market share of AVs by around 10% for a moderate annual reduction of 5% in AV price. In the same pricing scenario, 50% market share is forecasted to be achieved in 15.5 years in a 90% post-purchase satisfaction scenario,

but the same share would be attained in 17 years if 30% of early adopters are satisfied after buying AVs.

*5.3.2 Scenario 2: Effect of interpersonal social network*

To capture the impact of the interpersonal social network, we benchmark the forecasting results of ABM by integrating interdependent and independent consumer behavior models. The independent model does not (explicitly) account for the interpersonal network effects (i.e., $\delta = 0$). Even in the absence of network effects, consumer behavior model captures the direct impact of information sources on consumer uptake but fails to account for the indirect network effects transmitted through the structural equation of the WOM latent variable. The independent model also cannot capture the WOM dissemination by (dis)satisfied adopters (i.e., the model is insensitive to proportion of satisfied adopters). For interdependent ABM, the proportion of satisfied adopters is set to 90%. In this scenario, the initial AV price is set to $40,000 and is assumed to reduce annually by 5%. The forecasting results are displayed in Figure 9(b). The results indicate that the failure to account for social network effects can lead to a substantial underestimation of the market shares of the new technology. For instance, according to ABM with interdependent consumer behavior model, AV market share would be 50% in 15.5 years, but the same market share would be achieved in 18 years according to the independent model. These results corroborate our initial assertion that independent model may lead to unreliable forecasts due to its inability to capture the information propagation.

*5.3.3 Scenario 3: Effect of reduction in AV-involved crashes and related legal issues*

In this scenario, we quantify the effect of reduction in AV-involved crashes and the proportion of such crashes with no clear liability. As technology advances, AV-involved crashes and legal issues are likely to decrease, and the reduction is controlled in the ABM by curvature parameters (i.e., curvature crash and curvature legal) in Table 8. In particular, we consider four combinations of reductions in AV-involved crashes and legal issues regarding liability: (low, low), (low, high), (high, low), and (high, high) with corresponding curvatures (1.10, 5.00), (1.10, 1.00), (0.30, 5.00), and (0.30, 1.00), respectively. The AV-involved crashes and proportion of such crashes with legal issues in any year *t* can be obtained by plugging these values into the following function: $\text{Upper limit} - \left(\sqrt{\text{Upper limit}}\right) * \left(t/\text{Curvature}\right) - 1$ (as discussed in section 5.2). In this scenario, we set an initial AV price to $40,000, an annual reduction in AV price to 5%, and the proportion of satisfied adopters to 90%. Figure 9(c) shows the impact of all four possible combinations of curvatures on the adoption of AVs. The forecasting results show that the (high, high) scenario takes 14.5 years to achieve an AV market share of 50%, but the (low, low) scenario would attain the same market share in 16 years.

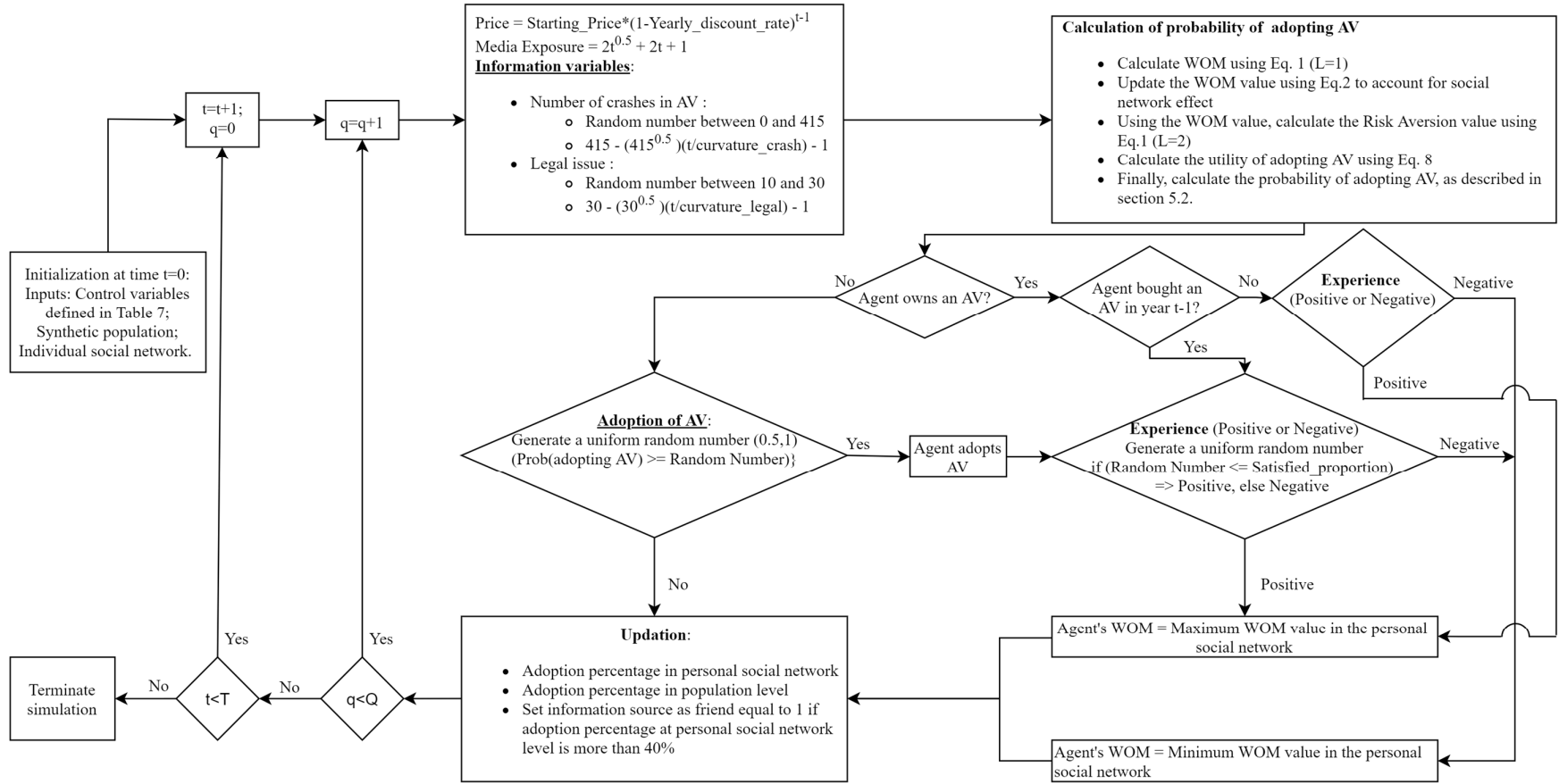

Figure 8: Agent-Based Simulation framework using Interdependent ICLV model.

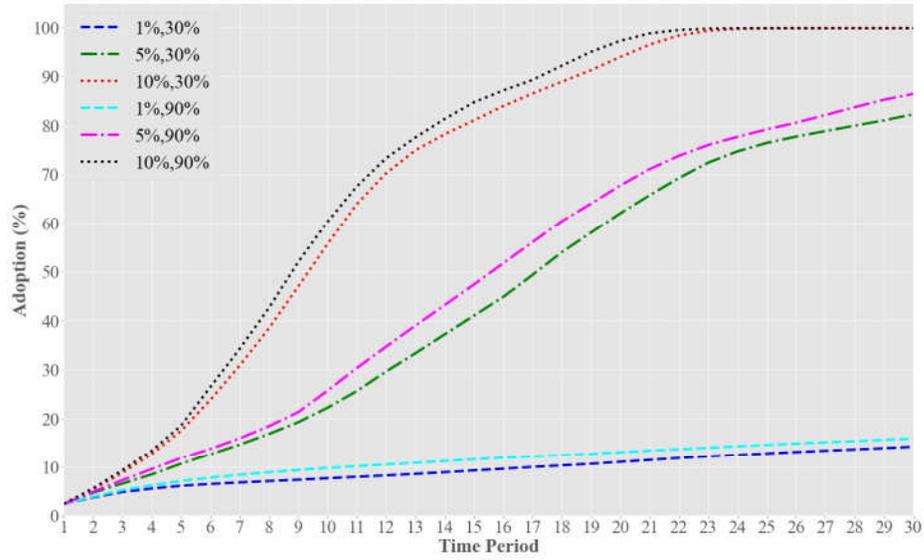

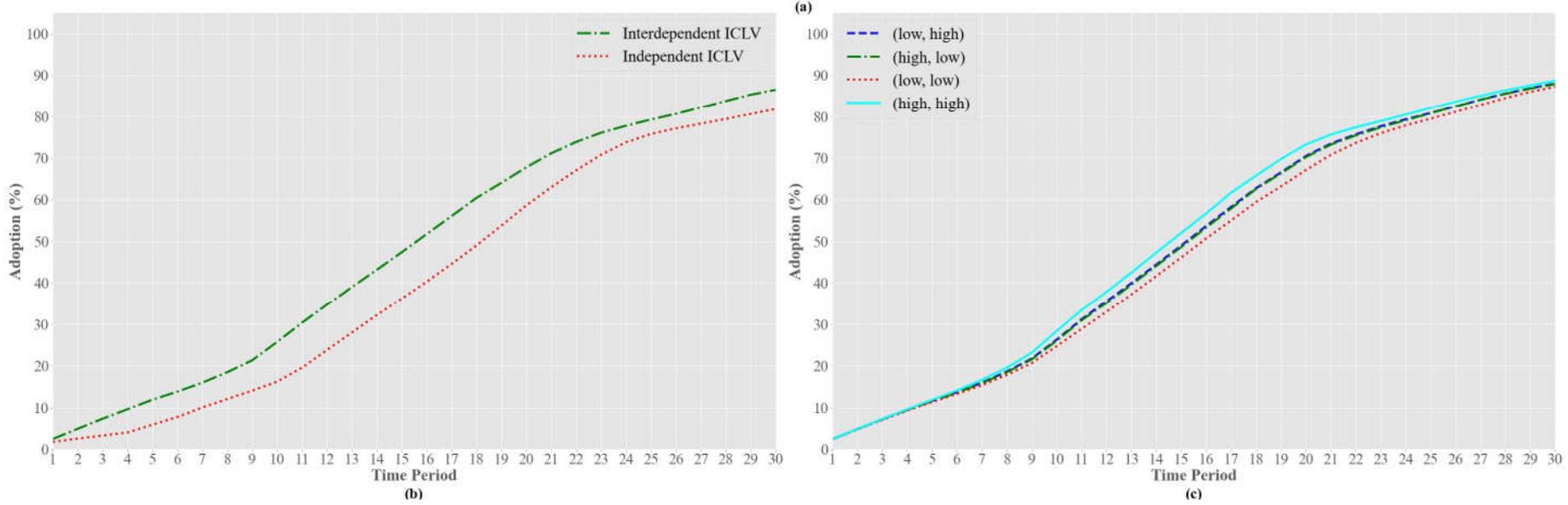

Figure 9: (a) Effect of AV price reduction and proportion of AV adopters with post-purchase satisfaction on AV adoption, (b) Effect of interpersonal social network on AV adoption (c) Effect of reduction in AV-involved crashes and proportion of such crashes with legal issues on AV adoption. (Weight matrix based on Gower distance with five social ties)

*5.3.4   AV adoption in Nashville*

We apply the ABM to assess the spatial (zip code-level) distribution of AV adoption levels in Nashville. We assume that AVs are introduced in 2021 with an initial price of $40,000. We also consider an annual reduction in AV price by 5% and the proportion of satisfied adopters as 75%. The forecasted AV adoption densities (the number of adopted AVs per square mile) in the years 2030, 2040 and 2050 for Gower distance based social network with five ties are shown in Figure 10. The plots indicate that, regardless of starting density, all zip codes exhibit marked improvement in AV density with time.

## 6. Conclusions and Future Work

Forecasting the adoption of the novel or "really new" products is of interest across multiple disciplines. Due to the lack of first-hand experience, potential early adopters of novel products actively expose themselves to word of mouth (WOM) information obtained from multiple sources (e.g., social networks and media) to assuage the potential risks. Not only do the existing consumer behavior models for novel products fail to capture the social network effects and interplay between WOM dissemination and risk aversion of consumers, but even cutting-edge forecasting models for such products also lack the integration of a calibrated consumer behavior model.

This study develops a general framework to forecast the adoption of novel products by combining a consumer behavior model with a population-level agent-based model (ABM). The consumer behavior is estimated using an integrated choice and latent variable (ICLV) model with WOM and risk aversion as two latent variables. The latent construct in ICLV captures spatial effects, i.e., the impact of homophilic peers in consumers' social networks. The discrete choice model in the ICLV accounts for the product's purchase price and adoption rates at individual's social network and city levels. We extend this ICLV model to account for the moderation of risk aversion behavior through WOM and panel effects in the choice component, and derive a maximum likelihood estimator to estimate the extended model. A calibrated consumer behavior model, synthetic population, and social network of all agents are embedded in an agent-based simulation framework to forecast the market share of the novel product. The proposed framework is applied to forecast the future adoption of autonomous vehicles (AV) in Nashville, USA over the next thirty years.

We calibrate the consumer behavior model with the data collected from an online stated preference survey of 1,495 residents from Nashville. The consumer behavior model identifies the risk-averse demographic groups and those with a tendency to spread the positive WOM about AVs. The results also indicate that reduction in AV-involved crashes and clear liability in these crashes is critical for positive WOM dissemination. Positive WOM turns out to be the driving factor in reducing the risk aversion of consumers. The results of the discrete choice component show that lower purchase price and higher social-network- and city-level adoption increase an individual's likelihood to purchase an AV. Whereas results of the consumer behavior model are new additions to the AV literature, the main contribution of this study stems from providing a general framework to quantify the effect of policy interventions on the adoption of AV technology.

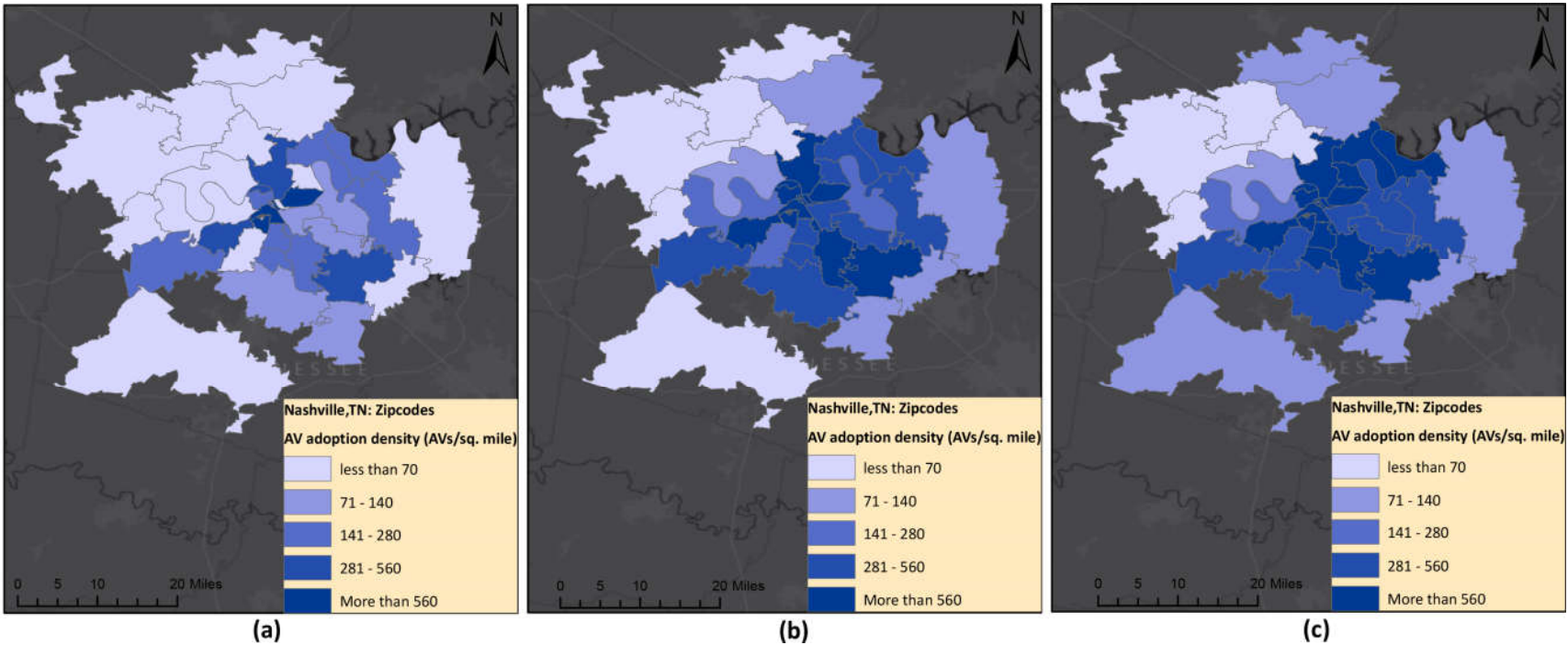

**Figure 10:** AV adoption density in Nashville, TN (a) 2030 (b) 2040 (c) 2050 (weight matrix based on Gower distance with five social ties).

The calibrated consumer model and synthetic population of Nashville is passed through the ABM to forecast the market share of AVs over the next 30 years (i.e., 2021-2050) in different scenarios. We quantify the effect of reduction in the purchase price, the extent of the post-purchase satisfaction, the importance of including social network effects, and the improvement in AV safety measures on the market share of AVs. In a moderate scenario – an annual price reduction of 5% at the initial price of $40,000, and 90% of adopters with post-purchase satisfaction, AVs are likely to attain 50% market share in 15.5 years and around 85% market share in 30 years after their market introduction.

As the study is not without limitations, it can be improved empirically and methodologically. First, we do not consider the rebuying or reselling of the AVs in the simulation framework. Future studies can develop an econometric model of vehicle ownership duration and the number of vehicles owned by the household. Second, we do not estimate the weight matrix in the ICLV estimation; instead, it is calculated based on the Euclidian distance or similarity in socio-demographic characteristics. Some studies in spatial statistics have estimated weight matrix elements, but those are not scalable (Bhattacharjee and Jensen-Butler, 2013; Qu and Lee, 2015). Future studies can build upon these studies to find ways to reduce the complexity by assuming zero effect (i.e., fixing the weight matrix element to zero) beyond a certain distance or the number of social ties. Third, we consider a fixed number of ties for every individual in the ICLV estimation. Potential ways to estimate the number of ties can be explored by subsequent studies.

Zhang, T., Gensler, S., Garcia, R., 2011. A study of the diffusion of alternative fuel vehicles: An agent-based modeling approach. J. Prod. Innov. Manag. 28, 152–168. https://doi.org/10.1111/j.1540-5885.2011.00789.x

Zhang, T., Tao, D., Qu, X., Zhang, X., Lin, R., Zhang, W., 2019. The roles of initial trust and perceived risk in public's acceptance of automated vehicles. Transp. Res. Part C Emerg. Technol. https://doi.org/10.1016/j.trc.2018.11.018

Zhao, Y., Joe, H., 2005. Composite likelihood estimation in multivariate data analysis. Can. J. Stat. 33, 335–356.

Zoellick, J.C., Kuhlmey, A., Schenk, L., Schindel, D., Blüher, S., 2019. Amused, accepted, and used? Attitudes and emotions towards automated vehicles, their relationships, and predictive value for usage intention. Transp. Res. Part F Traffic Psychol. Behav. 65, 68–78. https://doi.org/10.1016/J.TRF.2019.07.009
41

# Appendix

Note that Eq. 11 assumes that each individual is connected with everyone else in the network. However, in practice, spatial correlations are substantial enough to be considered for very few social ties. For instance, in this study, we create a spatial weight matrix considering only 5 or 10 social ties. Thus, we adopt the CML approach to write the likelihood for a pair of individuals $q$ and $q'$, which can be used to compute the likelihood for all pairs in the sample. The pair-wise likelihood is further aggregated to compute the sample likelihood.

To this end, we need to create additional selection matrices. We first create a selection matrix $\mathbf{D}_{qq'}$ of size $[2(H+T(I-1)) \times Q(H+T(I-1))]$ to extract the mean $(\bar{B}_{qq'})$ and covariance matrix $(\bar{\Omega}_{qq'})$ for the pair of individuals $q$ and $q'$, such that $\bar{B}_{qq'} = \mathbf{D}_{qq'}\bar{B}$, $\bar{\Omega}_{qq'} = \mathbf{D}_{qq'}\bar{\Omega}\mathbf{D}_{qq'}'$. To create $\mathbf{D}_{qq'}$, we create a matrix of zeros of the same size and insert two identity matrices of size $(H+T(I-1))$. The first identify matrix is inserted in the first $(H+T(I-1))$ rows and columns $[(q-1)*(H+T(I-1))+1]$ to $[q*(H+T(I-1))]$ of $\mathbf{D}_{qq'}$. The second identity matrix is insered in rows $(H+T(I-1)+1)$ to $[2(H+T(I-1))]$, and columns $[(q'-1)*(H+T(I-1))+1]$ to $[q'*(H+T(I-1))]$.

To explicitly write the *first term* (pairing of ordinal variables) of the CML expression in Eq. (11), we create a selection matrix $\mathbf{V}_{qq'}$ of size $(2H \times 2(H+T(I-1)))$. Specifically, we create a matrix of zeros of the same size as of $\mathbf{V}_{qq'}$ and insert two identity matrices of size $(H \times H)$. The first identity matrix is inserted in first $H$ rows and columns of $\mathbf{V}_{qq'}$ and insert another identity matrix in rows $(H+1)$ to $2H$ and columns $[H+T(I-1)+1]$ to $[2H+T(I-1)]$.

$\bar{B}_{V,qq'} = \mathbf{V}_{qq'}\bar{B}_{qq'}$, $\bar{\Omega}_{V,qq'} = \mathbf{V}_{qq'}\bar{\Omega}_{qq'}\mathbf{V}_{qq'}'$,

$\psi_{low} = \left(\psi'_{low}\left[\{(q-1)*H+1\}:\{(q-1)*H+H\}\right], \psi'_{low}\left[\{(q'-1)*H+1\}:\{(q'-1)*H+H\}\right]\right)'$,

$\psi_{up} = \left(\psi'_{up}\left[\{(q-1)*H+1\}:\{(q-1)*H+H\}\right], \psi'_{up}\left[\{(q'-1)*H+1\}:\{(q'-1)*H+H\}\right]\right)'$,

$\mu_{V,low} = \dfrac{\psi_{low} - \bar{B}_{V,qq'}}{\sqrt{diag(\bar{\Omega}_{V,qq'})}}$, $\mu_{V,up} = \dfrac{\psi_{up} - \bar{B}_{V,qq'}}{\sqrt{diag(\bar{\Omega}_{V,qq'})}}$

To explicitly write the *second term* (pairing of ordinal variables with nominal variables) of the CML expression in Eq. (11), we create a rearrangement matrix $\mathbf{\Delta}_{qq'}$ of size $[2(H+T(I-1)) \times 2(H+T(I-1))]$ that brings together ordinal and nominal responses together for simplicity. We create a matrix of zeros of the same size as of $\mathbf{\Delta}_{qq'}$ and insert four identity matrices, two of size $(H \times H)$ and the remaining two of size $[T(I-1) \times T(I-1)]$. The first identity matrix of size $(H \times H)$ is inserted in the first $H$ rows and columns, and the second identity matrix in rows $(H+1)$ to $2H$ and columns $[H+T(I-1)+1]$ to $[2H+T(I-1)]$ of $\Delta_{qq'}$. Similarly, the first identity matrix of size $[T(I-1) \times T(I-1)]$ in inserted in rows $(2H+1)$ to $[2H+T(I-1)]$ and columns $(H+1)$ to $[H+T(I-1)]$ and the second identity matrix in rows $[2H+T(I-1)+1]$ to $[2(H+T(I-1))]$ and columns $[2H+T(I-1)+1]$ to $2(H+T(I-1))$ of $\Delta_{qq'}$. A selection matrix $\mathbf{H}_{ht}$ of size $[I \times 2(H+T(I-1))]$ is also defined. To create $\mathbf{H}_{ht}$, we start with a matrix of zeros of the same size. Subsequently, the cell $(1,h)$ is filled with value of 1 and an identity matrix of size $(I-1) \times (I-1)$ is inserted in rows 2 to $I$ and columns $[2H+(t-1)(I-1)+1]$ through $[2H+t(I-1)]$. Then we define the following notations:

$$\bar{B}_{\Delta,qq'} = \Delta_{qq'}\bar{B}_{qq'}, \quad \bar{\Omega}_{\Delta,qq'} = \Delta_{qq'}\bar{\Omega}_{qq'}\Delta'_{qq'},$$
$$\psi^h_{low} = (\psi_{low}[h], \mathbf{0}_{I-1}), \quad \psi^h_{up} = (\psi_{up}[h], \mathbf{0}_{I-1}).$$

Finally, to write the *third term* (pairing of nominal variables) in Eq. 11 explicitly, we create a matrix $\mathbf{E}_{tt'}$ of size $[2(I-1) \times 2(H+T(I-1))]$. Specifically, we create a matrix of zeros of the same size as of $\mathbf{E}_{tt'}$ and insert two identity matrices of size $[(I-1) \times (I-1)]$. The first identity matrix is inserted in first $(I-1)$ rows and columns $[2H+(t-1)(I-1)+1]$ through $[2H+t(I-1)]$. The second identity matrix is inserted in the last $(I-1)$ rows and columns $[2H+(t'-1)(I-1)+1]$ through $[2H+t'(I-1)]$. The surrogate likelihood function a pair of individuals $q$ and $q'$ is:

$$L^{qq'}_{CML}(\Theta) = \left( \prod_{h=1}^{2H-1} \prod_{h'=h+1}^{2H} \begin{bmatrix} \Phi_2\left(\{\mu_{V,up}[h], \mu_{V,up}[h']\}, \bar{\Omega}^{hh'}_{V,qq'}\right) - \Phi_2\left(\{\mu_{V,up}[h], \mu_{V,low}[h']\}, \bar{\Omega}^{hh'}_{V,qq'}\right) \\ -\Phi_2\left(\{\mu_{V,low}[h], \mu_{V,up}[h']\}, \bar{\Omega}^{hh'}_{V,qq'}\right) + \Phi_2\left(\{\mu_{V,low}[h], \mu_{V,low}[h']\}, \bar{\Omega}^{hh'}_{V,qq'}\right) \end{bmatrix} \right) \times$$
$$\left( \prod_{h=1}^{2H} \prod_{t=1}^{2T} \Phi_I\left[(\psi^h_{up} - \mathbf{H}_{ht}\bar{B}_{\Delta,qq'}), \mathbf{H}_{ht}\bar{\Omega}_{\Delta,qq'}\mathbf{H}'_{ht}\right] - \Phi_I\left[(\psi^h_{low} - \mathbf{H}_{ht}\bar{B}_{\Delta,qq'}), \mathbf{H}_{ht}\bar{\Omega}_{\Delta,qq'}\mathbf{H}'_{ht}\right] \right) \times \quad (A.1)$$
$$\left( \prod_{t=1}^{2T-1} \prod_{t'=t+1}^{2T} \Phi_{2(I-1)}\left[-\mathbf{E}_{tt'}\bar{B}_{qq'}, \mathbf{E}_{tt'}\bar{\Omega}_{qq'}\mathbf{E}'_{tt'}\right] \right).$$

where $\bar{\Omega}^{hh'}_{V,qq'}$ is a $(2 \times 2)$ submatrix of $\bar{\Omega}_{V,qq'}$ corresponding to $h$ and $h'$ ordinal indicators, and $\Phi_{2(I-1)}\left[-\mathbf{E}_{tt'}\bar{B}_{qq'}, \mathbf{E}_{tt'}\bar{\Omega}_{qq'}\mathbf{E}'_{tt'}\right]$ is a cummulative distribution function of $2(I-1)$ dimensional MVN with mean $(-\mathbf{E}_{tt'}\bar{B}_{qq'})$ and covariance $(\mathbf{E}_{tt'}\bar{\Omega}_{qq'}\mathbf{E}'_{tt'})$.

The maximum dimension of integration in Eq. (A.1) is $2(I-1)$. We use GHK-simulator with quasi-random draws to evaluate MVNCDF accurately. Readers are referred to Train (2000) and Bhat (2003) for a detailed discussion on benefits of using quasi-random draws. The asymptotic covariance matrix of the model parameters is obtained by inverting

Godambe's (1960) sandwich information matrix, which requires Hessian and Jacobian matrices of the loglikelihood function at the convergence. The Jacobian matrix for models with spatial dependencies is computed using window sampling approach. Readers are referred to Zhao and Joe (2005), Bhat (2014), and Sidharthan and Bhat (2012) for more details on the calculations of the Hessian and the Jacobian matrix.

$\mathbf{M} = \text{zeros}\big(Q(H+T(I-1)), Q(H+TI)\big)$
**for** q in 1:Q
    row_ms = $(q-1)*(H+T(I-1))+1$
    row_me = $q*(H+T(I-1))$
    col_ms = $(q-1)*(H+TI)+1$
    col_me = $q*(H+TI)$
    $\mathbf{M}_q = \text{zeros}\big(H+T(I-1), H+TI\big)$
    $\mathbf{M}_q[1:H, 1:H] = \mathbf{IDEN}_H$
    Iden_mat = $\mathbf{1}_{I-1}$
    O_neg = -1*ones$(I-1,1)$
    **for** t in 1:T
        if($i_q^t == 1$)
            temp = O_neg~Iden_mat
        elseif($i_q^t == I$)
            temp = Iden_mat ~ O_neg
        else
            temp = Iden_mat[., $1:i_q^t-1$] ~ O_neg~Iden_mat[., $i_q^t:I-1$]
        end
        row_s = $H+(t-1)*(I-1)+1$
        row_e = $H+t*(I-1)$
        col_s = $H+(t-1)*I+1$
        col_e = $H+t*I$
        $\mathbf{M}_q[\text{row\_s}:\text{row\_e}, \text{col\_s}:\text{col\_e}]$ = temp
    $\mathbf{M}[\text{row\_ms}:\text{row\_me}, \text{col\_ms}:\text{col\_me}] = \mathbf{M}_q$

*Note*: where " ~ " refers to horizontal concatenation and $i_q^t$ is the chosen alternative at time period *t* by individual *q*.

**Algorithm A.1** An algorithm to generate **M** matrix.